\RequirePackage{fix-cm}

\documentclass[12pt,a4paper,natbib,fleqn,authoryear]{article}
\RequirePackage[OT1]{fontenc}
\long\def\symbolfootnote[#1]#2{\begingroup%
	\def\thefootnote{\fnsymbol{footnote}}\footnote[#1]{#2}\endgroup}
        
\usepackage{bm}
\usepackage{amsmath}
\usepackage{amssymb}
\usepackage{theorem}
\usepackage{xspace}
\usepackage{geometry}
\usepackage{array}
\usepackage{subfigure}
\usepackage{epsfig}
\usepackage{hhline}
\usepackage{bbm}
\usepackage{natbib}
 \usepackage{xcolor}
 \usepackage{float} 
 \usepackage{hyperref}

 \usepackage{booktabs}     
 \usepackage{multirow}     
 \usepackage{multicol}     
 \usepackage{makecell}     
 \usepackage{graphicx}     
 \usepackage{arydshln}     
 \usepackage{caption}

\usepackage[utf8]{inputenc}    
\DeclareMathAccent{\widehat}{\mathord}{largesymbols}{"62}
\DeclareMathAccent{\widetilde}{\mathord}{largesymbols}{"65}

\def\tr{\mathop{\mathrm{tr}}}

\def\eb*{\textrm{\mathversion{bold}$\mathbf{\beta^*}$\mathversion{normal}}}

\def\oo{\textrm{\mathversion{bold}$\mathbf{0}$\mathversion{normal}}}
\def\eb{\textrm{\mathversion{bold}$\mathbf{\beta}$\mathversion{normal}}}  
\def\ed{\textrm{\mathversion{bold}$\mathbf{\delta}$\mathversion{normal}}}

\def\eE{\mathbb{E}}

\def\e1{1\!\!1}

\theoremstyle{plain}

\newtheorem{theorem}{Theorem}[section]

\newtheorem{remark}{Remark}[section]

\newcommand{\beqn}{\begin{eqnarray*}}
\newcommand{\eeqn}{\end{eqnarray*}}
  
\def\ee1{\textrm{\mathversion{bold}$\mathbf{\varepsilon}$\mathversion{normal}}}

\newcommand{\N}{\mathbb{N}}
\newcommand{\R}{\mathbb{R}}
\newcommand{\PP}{\mathbb{P}}
\newcommand{\eSS}{\mathbb{S}}

\def\eX{\mathbf{X}}
\def\ex{\mathbf{x}}

\newcommand{\Var}{\mathbb{V}\mbox{ar}\,}

\def\argmin{\mathop{\mathrm{arg\,min}}} 
\def\oo{\textrm{\mathversion{bold}$\mathbf{0}$\mathversion{normal}}}

\begin{document}
\title {{\bf Model selection by cross-validation in an expectile  linear regression}}
 

 \author{Bilel BOUSSELMI$^1$ and Gabriela CIUPERCA$^{2}$
 	\date{}}

\maketitle
\setcounter{footnote}{1}
\footnotetext{\noindent \small{ESME Lyon, Lyon, France\\
		\textit{Email address}: bilel.bousselmi@esme.fr\\
		\indent $^2$Universit\'e de Lyon, Université Lyon 1, CNRS, UMR 5208, Institut Camille Jordan, Bat.  Braconnier, 43, blvd du 11 novembre 1918, F - 69622 Villeurbanne Cedex, France\\
		\textit{Email address}: Gabriela.Ciuperca@univ-lyon1.fr}}

 \begin{abstract}
 For  linear   models that may have asymmetric errors, we study variable selection by cross-validation.  The data are split into  training  and  validation sets, with the number of observations in the validation set  much larger than in the training set. For the model coefficients, the expectile or adaptive LASSO expectile estimators are calculated on the training set. These estimators will be used to calculate the cross-validation mean score (CVS)  on the validation set. We show that the model that minimizes CVS is consistent in two cases: when the number of explanatory variables is fixed or when it depends on the number of observations. Monte Carlo simulations confirm the theoretical results and demonstrate the superiority of our estimation method compared to two others in the literature.  The usefulness of the CV expectile model selection technique  is illustrated   by applying it to real data sets.
   \end{abstract}

\section{Introduction}
Cross-validation (CV) is one of the most widely used methods for model selection. It is valued for its simplicity and versatility in regression and classification frameworks. \cite{Stone.74} introduced the leave-one-out cross-validation technique which has enabled cross-validation methods to be developped over time for model selection. The subject has been the focus of extensive research in various studies. Readers are invited to consult the various cross-validation methods, with references, in the papers of \cite{Arlot.10} and \cite{Jhung.18}, for example. One of the cross-validation methods is the Train-Test split method.  This method also known as the hold-out method, is one of the earliest strategies proposed for evaluating predictive models. The idea of separating available data into two independent subsets: one for training and one for validation, emerged in the context of statistical learning and pattern recognition in the 1970s and 1980s (\cite{ Dev-Kit.82}; \cite{Ripley.96}). This approach provided a simple yet effective way to estimate how well a model generalizes to unseen data. Later, \cite{Kohavi.95} formalized the comparison between the hold-out method and more sophisticated resampling techniques such as cross-validation and bootstrap, highlighting both its conceptual simplicity and its sensitivity to the specific data split. The method has since been widely adopted as a baseline evaluation strategy, and remains an essential reference point in modern machine learning (see \cite{Hast-Tib-Fri.09} for more details). In conditional mean regression, \cite{Shao.93} provided the first theoretical justification of CV, showing that under appropriate conditions it achieves variable-selection consistency when based on the squared error loss. \cite{Zhang.93} extended the simple leave-one-out  cross-validation method to multifold CV.   A review on the cross validation and applications in ecology  are presented in \cite{Yates.23}. Concerning allowing  optimal sample size selection of the training/test set for fixed values of the total sample size, \cite{Afendras-Markatou.19} study the cases of repeated train–test CV and k-fold CV for certain decision rules. A discussion and references on CV can be found in the very interesting paper of \cite{Bates-Tibshirani.24}. Without being exhaustive let us mention a few more recent works on the CV procedure. \cite{Silva-Zanella.24} propose a robust novel mixture estimator to compute Bayesian leave-one-out cross-validation criteria. \cite{Fang-Yang-Tian.22} introduced a  cross-validation procedure to select the penalty factor in least squares model averaging that leads to an asymptotically optimal estimator, while \cite{Rabinowicz-Rosset.22} considered the CV for correlated data. \cite{Wieczorek-Leit.22} propose sufficient conditions for model selection consistency of forward selection with a data-driven stopping rule, based on a sequential variant of cross-validation. A flexible framework that facilitates multiple types of response for multi-penalty high-dimensional ridge regression is proposed in the paper of  \cite{van de Wiel-van Nee-Rauschenberger.21}.  More recently,  \cite{Kim-Lee-Shin-Jung.25}  proved that CV with the check quantile loss  consistently identifies the true model for both penalized and non-penalized estimators when the validation set is much larger than the training set. We will mainly compare our results with the theoretical and numerical results of the latter article. For more information on other aspects of cross-validation, readers may also refer to the papers by \cite{Wang.Cai.22}, \cite{Xiong-Shang.21}, \cite{Wei.Wang.21}, \cite{Algamal.20}, \cite{She-Tran.19}. \\
Expectile regression, introduced by \cite{Newey-Powell.87}, extends least squares estimation to asymmetric settings by minimizing a weighted quadratic loss. This makes expectiles particularly attractive when error distributions are skewed but possess finite moments. Expectile regression shares many features with quantile regression, such as robustness to distributional asymmetry and the ability to explore conditional distributions beyond the mean. However, unlike quantile regression, expectile regression leverages a differentiable loss function, which often leads to more efficient estimation and facilitates theoretical analysis. Despite these appealing properties, the theoretical role of CV in expectile regression (especially in the context of variable selection) remains largely unexplored.\\
The aim of this paper is to to study the CV in linear expectile regression models. We consider both fixed- and high-dimensional regimes, where the number of covariates may grow with the sample size. A crucial requirement is that the validation set must asymptotically dominate the training set in size, a condition that contrasts with the common practice of prioritizing large training samples. Then, based on the expectile or adaptive LASSO expectile  estimators found on the training set, we prove that CV consistently selects the true model under mild regularity assumptions. The theoretical results are confirmed by simulations, which also demonstrate the advantages of the proposed method compared to others described in the literature. The practical interest is demonstrated by applying the CV expectile method to two data sets. \\ 
The remainder of the paper is organized as follows. Section \ref{section_Model} introduces the model and the selection criterion by cross-validation based on expectile  and adaptive expectile LASSO estimation methods. For the three cases considered in respect to the number of model parameters, we show in Section \ref{section_result_theorique} the consistency of the models. Section \ref{section_simus} presents the simulation results while Section \ref{section_applications}  contains applications on real datasets. Section \ref{section_conclusion} summarizes the theoretical and numerical strengths of the CV expectile method, as well as the conclusions of the numerical analysis. The proofs of the theoretical results are relegated in Section \ref{section_proofs}.

\section{Model and methods}
\label{section_Model}
Let us start with some notation that will be used throughout the rest of the paper. All vectors and matrices are denoted by bold symbols and all vectors are column.  For a vector or matrix, the symbol $T$ at the top right is used for its transpose. For a vector  $\textbf{v}$, we denote   by $\|\textbf{v}\|_1$ its $L_1$-norm, $\|\textbf{v}\|_2$ its euclidean norm and by  $\|\textbf{v}\|_\infty$ its $L_\infty$-norm. Moreover, for a positive definite matrix, we denote by $\mu_{min}(.)$ and $\mu_{max}(.)$ its largest and smallest eigenvalues.  If $E$ is an event, $\e1_E$ denotes the indicator function that the event $E$ occurs.  Given a set ${\cal S}$, we denote its cardinality by $|{\cal S}|$.  If $(a_n)_{n \in \N }$, $(b_n)_{n \in \N}$ are two positive deterministic sequences we use the notation  $a_n \gg b_n$, when $\lim_{n \rightarrow \infty} a_n/b_n=\infty$. The convergences in distribution, in probability and almost sure  as $n$ converges to infinity are denoted by  $\overset{\cal L} {\underset{n \rightarrow \infty}{\longrightarrow}}$, $\overset{\PP} {\underset{n \rightarrow \infty}{\longrightarrow}}$, $\overset{a.s.} {\underset{n \rightarrow \infty}{\longrightarrow}}$, respectively. Throughout this paper, $C$ will always denote a generic constant, not depending on the size $n$ and its value is not of interest.\\

Let us consider the following linear model:  
\begin{equation}
\label{eq1}
Y_i=\eX_i^t \eb+\varepsilon_i, \qquad i=1, \cdots , n,
\end{equation}
with the parameter vector  $\eb \in \R^p$   and $\eb^*=(\beta^*_1, \cdots , \beta^*_p)$ its true value (unknown). The dimension of $\eb^*$ may depend on $n$, but the values of $\beta^*_j$ does not depend on $n$, for all  $j=1, \cdots , p$. The vector $\eX_i$ contains the values of the  $p$  explanatory deterministic variables and $Y_i$ the values of response variable for observation $i$. The values $(Y_i, \eX_i)$ are known for $i=1, \cdots , n$ and the model errors $(\varepsilon_i)_{1 \leqslant i \leqslant n}$ are random.\\
Let the index set of the non null parameters, 
\[
{\cal M}^* \equiv \big\{   j \in \{ 1, \cdots, p\}; \, \beta^*_j \neq 0\big\}.
\]
Since $\eb^*$ is unknown then the set ${\cal M}^*$ is also unknown.  \\

For a real fixed $\tau \in (0,1)$, let the expectile function $\rho_\tau(.)$ be defined as 
\[\rho_\tau(x)=|\tau - \e1_{x <0}|x^2, \qquad \textrm{with} \quad x \in \R.
\]
Consider the functions:
$
g_\tau(x) \equiv \rho'_\tau(x-t) |_{t=0}=2 \tau x \e1_{x \geq 0}+2(1-\tau)x \e1_{x<0}$, $\quad h_\tau(x) \equiv \rho''_\tau(x-t)_{t=0}=2 \tau \e1_{x \geq 0}+2(1-\tau) \e1_{x<0}$. \\
Note that the expectile estimation method  introduced by \cite{Newey-Powell.87} is used when the moments of $\varepsilon$ exist, but its distribution is asymmetric. If $\tau = 0.5$, we recover the classical least squares method.  \\

The model errors $(\varepsilon_i)_{1 \leqslant i \leqslant n}$ of model (\ref{eq1}) satisfy the following condition:
\begin{description}
\item  \textbf{(A1)} $(\varepsilon_i)_{1 \leqslant i \leqslant n}$ are i.i.d. such that $\eE[\varepsilon^2_i]< \infty$ and the $\tau$-th expectile of $\varepsilon$ is zero: $\eE[g_\tau(\varepsilon)]=0$.
\end{description}

\noindent The deterministic design $(\eX_i)_{1 \leqslant i \leqslant n}$ satisfies the assumption:
\begin{description}
\item \textbf{(A2)} There exist two constants $m_0, M_0$ such that $0 < m_0 \leq \mu_{min}\big(n^{-1} \sum^n_{i=1} \eX_i \eX_i^t\big)\leq \mu_{max}\big(n^{-1} \sum^n_{i=1} \eX_i \eX_i^t\big) \leq M_0 < \infty$.
\end{description}
To a set of indices ${\cal M} \subseteq \{1, \dots, p\}$, we associate a model of the form:
\[
Y_i=\eX^\top_{i,{\cal M}} \eb_{\cal M}+\varepsilon_i, \qquad i=1, \cdots, n,
\]
with $\eb_{\cal M}$ a subvector of $\eb \in \R^p$ corresponding to the indices in ${\cal M}$ and $\eX_{i,{\cal M}}$  the corresponding subvector of $\eX_i$. For simplicity, we refer to ${\cal M}$ as a model.\\
A model ${\cal M}_o \subseteq \{1, \cdots, p\}$ is over-specified if ${\cal M}^* \subsetneq {\cal M}_o $  and a model ${\cal M}_u$ is under-specified when ${\cal M}^*  \not \subset {\cal M}_u$.\\
We also need an identifiability assumption for the  estimated model using the expectile method. For any submodel ${\cal M}_u$ such that ${\cal M}^* \subsetneq {\cal M}_u$, we consider the following assumption:
\begin{description}
	\item {\bf{(A3)}} 
$ \liminf_{n \rightarrow\infty} n^{-1} \sum^n_{i=1} \big(\rho_\tau(Y_i-\eX_{i,{\cal M}_u}^\top\eb^*_{{\cal M}_u}) -\rho_\tau(\varepsilon_i)\big)=\liminf_{n \rightarrow\infty}  k_n>0.$ 
\end{description}
Assumption (A1) is commonly required for expectile models (see for example \cite{Newey-Powell.87}, \cite{Ciuperca.21}, \cite{Gu-Zou.16}, \cite{Liao-Park-Choi.18}, \cite{Zhao-Chen-Zhang.18}). Assumption (A2) is standard in a linear model and necessary for parameter identifiability (see \cite{Ciuperca.21}, \cite{Liao-Park-Choi.18}). \\

To better understand  assumption (A3), let us take $\tau =1/2$ if ${\cal M}^* \subsetneq {\cal M}_u$. Then the right-side in assumption {(A3)} becomes $$n^{-1} \sum^n_{i=1} \big( \eX^\top_{i,{\cal M}^* \setminus {\cal M}_u} \eb^*_{{\cal M}^* \setminus {\cal M}_u} \big)^2 - 2 n^{-1}\sum^n_{i=1} \varepsilon_i \eX^\top_{i,{\cal M}^* \setminus {\cal M}_u} \eb^*_{{\cal M}^* \setminus {\cal M}_u}.$$ On the other hand, by the law of large numbers,  using assumption (A1), we have:  $$ n^{-1}\sum^n_{i=1} \varepsilon_i \eX^\top_{i,{\cal M}^* \setminus {\cal M}_u} \eb^*_{{\cal M}^* \setminus {\cal M}_u} \overset{a.s.} {\underset{n \rightarrow \infty}{\longrightarrow}} 0.$$
Then, applying assumption (A2), assumption (A3) becomes: 
\begin{align*}
n^{-1} \sum^n_{i=1} \big( \eX^\top_{i,{\cal M}^* \setminus {\cal M}_u} \eb^*_{{\cal M}^* \setminus {\cal M}_u} \big)^2& = {\eb^*}^\top_{{\cal M}^* \setminus {\cal M}_u} n^{-1} \sum^n_{i=1} \eX_{i,{\cal M}^* \setminus {\cal M}_u} \eX^\top_{i,{\cal M}^* \setminus {\cal M}_u} \eb^*_{{\cal M}^* \setminus {\cal M}_u}\\
& {\underset{n \rightarrow \infty}{\longrightarrow}} {\eb^*}^\top_{{\cal M}^* \setminus {\cal M}_u} \lim_{n  \rightarrow \infty} \bigg(n^{-1} \sum^n_{i=1} \eX_{i,{\cal M}^* \setminus {\cal M}_u} \eX^\top_{i,{\cal M}^* \setminus {\cal M}_u} \bigg)\eb^*_{{\cal M}^* \setminus {\cal M}_u},
\end{align*}
which is consistent with the definition of ${\cal M}^*$ for the LS model. Then, assumption (A3)  is an  identifiability condition for the expectile method, which is a generalization of LS. \\

Additional assumptions on the design will be formulated  in the following two sub-sections, depending on the dimension $p$ as function of $n$.\\

Based on the definition of the Train-test split CV method, we consider a training set ${\cal A} \subsetneq \{1, \cdots n \}$ chosen randomly and containing $n_{\cal A} = |{\cal A}|$ observations with $n_{\cal A}$ depending on $n$. The validation (test) set is ${\cal V} \equiv \{1, \cdots, n\} \setminus {\cal A}$, which has $n_{\cal V} \equiv |{\cal V}| = n - n_{\cal A}$ observations.\\

The proposed model selection criterion in the present paper will be examined in cases where the number of validation observations is much larger than the number of training observations. This assumes the following:
\begin{description}
	\item 
 \textbf{(A4)} $n_{\cal A}  {\underset{n \rightarrow \infty}{\longrightarrow}} \infty$ and  $n_{\cal V} / n_{\cal A} {\underset{n \rightarrow \infty}{\longrightarrow}}  \infty$.
\end{description}
Note that the same assumption was considered in \cite{Kim-Lee-Shin-Jung.25}'s paper to select  by cross-validation the relevant explanatory variables of a quantile model. \\
 
So, let us now introduce the model selection criterion based on the expectile estimator calculated on the training set and on a   cross-validation mean calculated on the validation set. Therefore, for a model ${\cal M}$,     we first calculate on the training set ${\cal A}$ of size $n_{\cal A}$ the expectile estimator  $\widehat{\eb}_{n_{\cal A},{\cal M}}$ of $\eb_{\cal M}$:
\[
\widehat{\eb}_{n_{\cal A},{\cal M}} \equiv \argmin_{\eb \in \R^{|{\cal M}|}} \sum_{i\in {\cal A}} \rho_\tau (Y_i -\eX^\top_{i,{\cal M}} \eb_{\cal M}).
\]
The components of the expectile estimator  $\widehat \eb_{n_{\cal A},{\cal M}}$ are denoted by $\big(\widehat \beta_{(n_{\cal A},{\cal M});1}, \widehat \beta_{(n_{\cal A},{\cal M});2}, \cdots,$ $ \widehat \beta_{(n_{\cal A},{\cal M});p}\big)$.
After which, based on this estimator, we can compute the prediction of $Y_i$ on the validation observations, more precisely: $\widehat Y_i= \eX_{i,{\cal M}}^\top \widehat \eb_{n_{\cal A},{\cal M}}$, for $i \in {\cal V}$. These predictions allow us to calculate the residuals on the test set and thus define the cross-validation mean score corresponding to the expectile framework: 
\begin{equation}
	\label{eq2}
 CVS({\cal M})   \equiv \frac{1}{n_{\cal V}} \sum_{i \in {\cal V} }\rho_\tau (Y_i -\eX^\top_{i,{\cal M}} \widehat \eb_{n_{\cal A},{\cal M}}).
\end{equation}
Obviously, minimizing $CVS({\cal M})$ of relation (2) in respect to ${\cal M}$ amounts to minimizing the difference:
\begin{equation}
	\label{eq3i}
 CVS({\cal M}) -CVS({\cal M}^*) =\frac{1}{n_{\cal V}} \sum_{i \in {\cal V} }\rho_\tau (Y_i -\eX^\top_{i,{\cal M}} \widehat \eb_{n_{\cal A},{\cal M}}) - \frac{1}{n_{\cal V}} \sum_{i \in {\cal V} }\rho_\tau (Y_i -\eX^\top_{i,{\cal M}^*} \widehat \eb_{n_{\cal A},{\cal M}^*})
\end{equation}
The criterion proposed in this paper for estimating model ${\cal M}^*$ is as follows:   
\begin{equation}
\label{eq3}
\widehat {\cal M}_n \equiv \argmin_{{\cal M} \subseteq \{1, \cdots, p\} } CVS({\cal M}).
\end{equation}
 We call this estimator the CV expectile  estimator.\\
Before examining this criterion in the following two subsections, whether $p$ is fixed or dependent on $n$, let us consider $r_{{n}_{\cal A}}$  be a deterministic sequence that diverges to $\infty$ as $n \rightarrow \infty$, which depends on the number of observations $n_{\cal A}$ and on the number $p$ of parameters. In fact, $r_{n_{\cal A}}^{-1}$ is the convergence rate of $\widehat \eb_{n_{\cal A},{\cal M}}$ towards $\eb^*_{\cal M}$ for a given model ${\cal M}$. \\
Finally, for a model ${\cal M}$ estimated on the training observations ${\cal A}$, we consider the $|{\cal M}|$-dimensional random vector $\widehat \ed_{n_{\cal A},{\cal M}} \equiv r_{n_{\cal A}} \big(\widehat \eb_{n_{\cal A},{\cal M}}- \eb^*_{\cal M} \big)$.\\

We can also propose a model selection criterion  based on a penalized estimation method on the training set. To do this, let us introduce the following stochastic process with adaptive LASSO penalty:
\[
R_{n_{\cal A},{\cal M}}(\eb)\equiv \sum^{n_{\cal A}}_{i=1} \rho_\tau (Y_i - \eX_{i,{\cal M}}^t \eb_{\cal M})+n_{\cal A} \lambda_{n_{\cal A}} \sum^p_{j=1} \widehat \omega_{(n_{\cal A},{\cal M}),j} |\beta_j|,
\] 
where
\[
  \widehat \omega_{(n_{\cal A},{\cal M}),j}  =|\widehat \beta_{(n_{\cal A},{\cal M}),j}|^{-\gamma}, \quad j =1, \cdots , p
\]
 and  $\gamma \in (0,1]$ a known constant. \\
This last random  process is used to propose the adaptive LASSO expectile  estimator  that automatically  detects the zero and non-zero parameters of  $\eb$ (see \cite{Liao-Park-Choi.18}, \cite{Ciuperca.21}),  defined by:
\begin{equation}
	\label{Bal}
\widehat \eb_{n_{\cal A},{\cal M}}^{(aLASSO)} \equiv \argmin_{\eb \in \R} R_{n_{\cal A},{\cal M}}(\eb).
\end{equation}
The properties of the estimator $\widehat \eb_{n_{\cal A},{\cal M}}^{(aLASSO)}$ were studied by \cite{Liao-Park-Choi.18} for fixed $p$.  Its  convergence rate  towards $\eb^*$ is of order $n_{\cal A}^{-1/2}$ and  $\widehat \eb_{n_{\cal A},{\cal M}}^{(aLASSO)}$ satisfies the oracle properties, that is: automatic selection of significant variables with probability converging to $1$ and asymptotic normality for the non-zero estimators. If $p$ depends on $n$, the estimators $\widehat{\eb}_{n_{\cal A},{\cal M}}$ and $ \widehat \eb_{n_{\cal A},{\cal M}}^{(aLASSO)} $ were studied by \cite{Ciuperca.21}.  \\
Then, the cross-validation mean score  calculated for the adaptive LASSO expectile estimator is the following:
\begin{equation}
a CVS({\cal M})   \equiv \frac{1}{n_{\cal V}} \sum_{i \in {\cal V} }\rho_\tau (Y_i -\eX^\top_{i,{\cal M}} \widehat \eb^{(aLASSO)}_{n_{\cal A},{\cal M}})
\label{ta}
\end{equation}
and the corresponding model estimator:
\begin{equation}
	\label{Man}
\widehat {\cal M}^{(a)}_n \equiv \argmin_{{\cal M} \subseteq \{1, \cdots, p\} } aCVS({\cal M}).
\end{equation}
We will call $\widehat {\cal M}^{(a)}_n$ the CV adaptive LASSO expectile model estimator.\\
The consistency of the model estimator $\widehat {\cal M}_n$ and $\widehat {\cal M}^{(a)}_n$ will be studied in the following  section, while the proofs of the main results are postponed in Section \ref{section_proofs}.\\
The question that arises is: what is the point of using the cross-validation mean score criterion to select the model if the adaptive LASSO expectile method already does so? Another question that arises is why use the aCVS criterion of (\ref{Man}) and not the CVS? The answers to these questions can be found in Section \ref{section_simus}, where simulation results will be presented.\\
The model estimators $\widehat {\cal M}_n$ and $\widehat {\cal M}^{(a)}_n$  will be studied numerically in Section \ref{section_simus}, the simulations confirming their consistency.
\section{Asymptotic behavior of CV model estimators}
\label{section_result_theorique}
In this section, we study the consistency of the CV expectile estimators, first in the case where $p$ is fixed and then when it depends on $n$.
\subsection{Case of fixed $p$}
\label{subsection_pfixed}
In this subsection, we study the asymptotic behavior of  the model estimators $\widehat {\cal M}_n$ defined by (\ref{eq3}) and $\widehat {\cal M}^{(a)}_n$ defined by (\ref{Man}), when  the dimension $p$ does not depend on $n$.
In this case, we consider $r_{n_{\cal A}}= n_{\cal A}^{1/2}$.\\
To study $\widehat {\cal M}_n $, we need to know the asymptotic behavior of the expectile estimator $\widehat \eb_{n_{\cal A},{\cal M}}$ of $\eb$  for a given model ${\cal M}$.
 Theorem 2.1(i) of \cite{Ciuperca.21} implies  that for  a given model ${\cal M}$, under assumptions (A1) and (A2),   the expectile estimator is consistent, that is $\widehat \eb_{n_{\cal A},{\cal M}} \overset{\PP} {\underset{n \rightarrow \infty}{\longrightarrow}}  \eb^*_{\cal M}$, with a convergence rate of order $r^{-1}_{n_{\cal A}}=n_{\cal A}^{-1/2}$. Then, we deduce that  $\widehat \ed_{n_{\cal A},{\cal M}} =O_\PP(1)$, for all ${\cal M}\subseteq \{1, \cdots, p\}$. With this convergence rate, using a similar approach to the proof of   Theorem 1 in \cite{Liao-Park-Choi.18}, it can be shown that  under the same assumptions (A1), (A2), the random vector $\widehat \ed_{n_{\cal A},{\cal M}}$ is asymptotically normal, i.e. the convergence $\widehat \ed_{n_{\cal A},{\cal M}} \overset{\cal L} {\underset{n \rightarrow \infty}{\longrightarrow}} {\cal N}(\oo_{|{\cal M}|}, ...) \equiv \ed_{\cal M}$ occurs.\\

After these introductory remarks, we can demonstrate by the following theorem the consistency of the model $\widehat {\cal M}_n$ defined by (\ref{eq3}) .
\begin{theorem}
	\label{Theorem_pfix}
Under assumptions (A1), (A2), (A3), (A4),  we have: $\PP[\widehat {\cal M}_n ={\cal M}^*]{\underset{n \rightarrow \infty}{\longrightarrow}}  1$.
\end{theorem}

At the beginning of this section, we saw that we can construct the adaptive LASSO expectile estimator based on the expectile estimator. This estimator is calculated using relation (\ref{Bal}) on the training set and has the property of automatically selecting the true model.  The cross-validation mean score can be calculated using relation (\ref{ta}) based on the model's predictions made on the training set using $\widehat \eb_{n_{\cal A},{\cal M}}^{(aLASSO)} $.\\
Using the same argument as in \cite{Kim-Lee-Shin-Jung.25} and taking into account the sparsity of the adaptive  LASSO expectile estimator, as established by Theorem 3 in \cite{Liao-Park-Choi.18}, we can prove that Theorem \ref{Theorem_pfix} also holds for model estimator $\widehat {\cal M}^{(a)}_n$ defined by (\ref{Man}). 

\subsection{Case $p=p_n$, $p_n=O(n_{\cal A}^c)$}
\label{subsection_pn}
Consider now that the number $p$ of the explanatory variables depends on $n$, more precisely on the size $n_{\cal A}$ of the training set. We suppose that the training set ${\cal A}$ and $p$ are such that $p=p_n=O(n_{\cal A}^c)$, with the constant $c \in (0,1]$ and $p<n_{\cal A}$.  An example of this type of dependency would be $p=p_n=n^u$, with $u$ a positive constant. Taking into account assumption (A4), the constant $u$ is then strictly greater than  $c$.\\
Note that the cardinality $|{\cal M}^*|$ may itself depend on $n$.  An additional assumption to (A2) on the design must be considered:
\begin{description}
	\item \textbf{(A5)}   $\big( p_n/n_{\cal A}\big)^{1/2} \max_{1 \leqslant i \leqslant n} \| \eX_i \|_2 {\underset{n \rightarrow \infty}{\longrightarrow}} 0$,
	with the remark that $p_n/n_{\cal A} {\underset{n \rightarrow \infty}{\longrightarrow}} 0$.
\end{description}
Assumption (A5) was considered in \cite{Ciuperca.21}, where comments on it can be found. \\

\noindent On the other hand, for $i \in {\cal A}$, we have, $\tr(\eX_i^\top \eX_i)=\tr(\eX_i \eX_i^\top)=\|\eX_i\|^2_2 \leq \max_{1 \leqslant j \leqslant n_{\cal A}} \| \eX_j\|^2_2$ and taking into account assumption (A2) we obtain $ \max_{1 \leqslant j \leqslant n_{\cal A}} \| \eX_j\|^2_2 \geq Cp$. This last relation together with assumption (A5) imply $p_n n^{-1/2}_{\cal A} \rightarrow 0$ as $n \rightarrow \infty$ from where we deduce that if condition $p_n=O(n^c_{\cal A})$ and assumption (A5) hold then $c < 1/2$. So, we will deal with the two cases $0 \leq c < 1/2$ and $1/2 \leq c < 1$ separately.   \\
In the following two subsections, depending on whether  $c$ is less than or greater than 1/2, we present specific assumptions and demonstrate the consistency of the model estimator $\widehat {\cal M}_n$.  
\subsubsection{Case  $0<c<1/2$}
In this case, according to Theorem 2.1 in \cite{Ciuperca.21}, we consider as sequence $(r_{n_{\cal A}})_{n_{\cal A} \in \N}$ the following: $r_{n_{\cal A}}=n_{\cal A}^{1/2} p_n^{-1/2}$.\\
The fact that $p$ depends on $n$ requires the introduction of other assumptions   on $(\eX_i)_{i \in {\cal V}}$ in addition to (A2) and (A5):
\begin{description}
	\item \textbf{(A6)} $ n^{-1/2}_{\cal V}  \max_{j \in {\cal V}}\| \eX_j \|^2_2  {\underset{n \rightarrow \infty}{\longrightarrow}} 0$.
	\item \textbf{(A7)} $ r^{-1}_{n_{\cal A}}|{\cal M}^*|  \max_{i \in {\cal V}} \big\|\eX_{i,{\cal M}^*}\big\|^2_2  {\underset{n \rightarrow \infty}{\longrightarrow}} 0$. 
\end{description}
An example of when assumption (A6) holds if (A5) is satisfied is when $n^{-1/2}_{\cal V} n_{\cal A}/p_n $  is bounded for  sufficiently large $n$. Indeed, one can write:
\[
n^{-1/2}_{\cal V}\max_{j \in {\cal V}}\| \eX_j \|^2_2 \leq n^{-1/2}_{\cal V}\max_{1 \leqslant j \leqslant n}\| \eX_j \|^2_2= \bigg(\frac{p_n}{n_{\cal A}} \max_{1 \leqslant j \leqslant n}\| \eX_j \|^2_2\bigg) \frac{n_{\cal A}}{p_n} \frac{1}{n^{1/2}_{\cal V}}
\]
and taking into account assumption (A5), if $n^{-1/2}_{\cal V} n_{\cal A}/p_n$  is bounded then assumption (A6) is satisfied. Remark also that Assumption (A7)  follows from (A5) when $|{\cal M}^*|$ is bounded.\\
With these assumptions, we can prove the consistency of the CV expectile model estimator.
\begin{theorem}
	\label{Theorem_pn1}
	Under assumptions  (A1)-(A7), we have: $\lim_{n \rightarrow \infty}\PP[\widehat {\cal M}_n ={\cal M}^*]=  1$.
\end{theorem}
	\subsubsection{Case  $1/2\leq c \leq 1$}
	Assumptions (A2) and (A5) imply that $c<1/2$. If $p_n <n_{\cal A}$ and  $c \in [1/2,1]$, then assumption (A5) will need to be replaced with a different one:
\begin{description}
	\item {\bf  (A8)}  $\max_{1 \leqslant i \leqslant n} \| \eX_i\|_\infty < \infty$,
\end{description}
assumption considered, for example, in  \cite{Ciuperca.21}.\\

 In this case, the sequence $(r_{n_{\cal A}})_{n_{\cal A} \in \N}$ is $r_{n_{\cal A}}=a^{-1}_{n_{\cal A}}$, where $(a_n)_{n \in \N}$ is a deterministic sequence that converge to zero when $n \rightarrow \infty$ and $n^{1/2} a_n {\underset{n \rightarrow \infty}{\longrightarrow}} \infty$. In the proofs of theoretical results, the Hölder's inequality $|\eX_{i,{\cal M}} \eb_{\cal M}| \leq \| \eX_{i,{\cal M}}\|_\infty \|\eb_{\cal M} \|_1$ is used.
Since $\|.\|_\infty \leq \| . \|_2$,  assumptions (A6), (A7)  imply:
\begin{description}
	\item \textbf{(A6bis)} $ n^{-1/2}_{\cal V}  \max_{j \in {\cal V}}\| \eX_j \|^2_\infty  {\underset{n \rightarrow \infty}{\longrightarrow}} 0$.
	\item \textbf{(A7bis)} $r^{-1}_{n_{\cal A}}|{\cal M}^*|  \max_{i \in {\cal V}} \big\|\eX_{i,{\cal M}^*}\big\|^2_\infty  {\underset{n \rightarrow \infty}{\longrightarrow}} 0$. 
\end{description}
Assumption (A8) will replace (A5). However, instead of assumptions (A6) and (A7), lower  assumptions (A6bis) and (A7bis) will be considered, when $c \in [1/2,1]$.\\

Let us emphasize that in the case $c \in [1/2,1]$, for a model ${\cal M}$, the weight in the adaptive LASSO penalty calculated on the training set ${\cal A}$ is of the form:
\[
\widehat \omega_{({n_{\cal A},{\cal M}}),j}=\min\big(|\widetilde \beta_{({n_{\cal A},{\cal M}}),j}|^{-\gamma}, n_{\cal A}^{1/2}\big), \qquad \textrm{for } j\in \{1, \cdots, p\},
\]
where $\widetilde \eb_{n_{\cal A},{\cal M}}$  is either the expectile estimator $\widehat \eb_{n_{\cal A},{\cal M}}$  or  the   LASSO expectile estimator,  the latter being the minimiser of:
\[
n_{\cal A}^{-1} \sum^{n_{\cal A}}_{i=1}\rho_\tau (Y_i - \eX_{i,{\cal M}} \eb_{\cal M}) + \nu_{n_{\cal A}} \| \eb_{\cal M} \|_1,
\]
with $\nu_{n_{\cal A}} $ a strictly positive deterministic sequence such that $\nu_{n_{\cal A}}  {\underset{n \rightarrow \infty}{\longrightarrow}} 0$. \cite{Ciuperca.21} showed that, concerning the consistency of the adaptive LASSO expectile estimator, if $\lim_{n_{\cal A}  \rightarrow \infty}\lambda_{n_{\cal A}} |{\cal M}^*|^{1/2} b^{-1}_{n_{\cal A}}  = 0$ and  assumptions (A1), (A2), (A8) are met, then $\| \widehat \eb^{(aLASSO)}_{n_{\cal A}, {\cal M}} - \eb^*_{\cal M}\|_1=O_\PP(b_n)$, where $b_n$ is a deterministic sequence converging to 0 when $n \rightarrow \infty$.\\

With these elements, the consistency of the model estimator $\widehat {\cal M}_n$   can be shown by the following theorem.

\begin{theorem}
	\label{Theorem_pn2}
	Under assumptions  (A1)-(A4), (A6bis), (A7bis), (A8) we have:  
	$$\lim_{n \rightarrow \infty}\PP[\widehat {\cal M}_n ={\cal M}^*]=  1.$$
\end{theorem}
\begin{remark}In both cases of $c$, when $p=p_n=n^c$, it can be shown, in a similar manner to the case of fixed $p$, that model selection is achieved by employing the cross-validation criterion (\ref{Man}) for the adaptive  LASSO expectile estimator.
\end{remark}
\begin{remark}
	In  paper  \cite{Kim-Lee-Shin-Jung.25}, the unpenalized case when $p_n \rightarrow \infty$ is not considered. Furthermore, in the context of the penalised case, when $p_n \rightarrow \infty$, only the SCAD penalty is considered, under the very strong conditions, particularly condition (D3): $ {p^3_{n_{\cal A}}}/{n_{\cal A}} \rightarrow \infty$. When $|{\cal M}^*|$ diverges, an even stronger condition is imposed in \cite{Kim-Lee-Shin-Jung.25}: $|{\cal M}^*|^3 p^3_n/n  \rightarrow \infty$.
\end{remark}

\section{Simulation studies}
\label{section_simus}
\begin{figure}
	\begin{tabular}{cc}
			\includegraphics[width=0.45\linewidth,height=3.5cm]{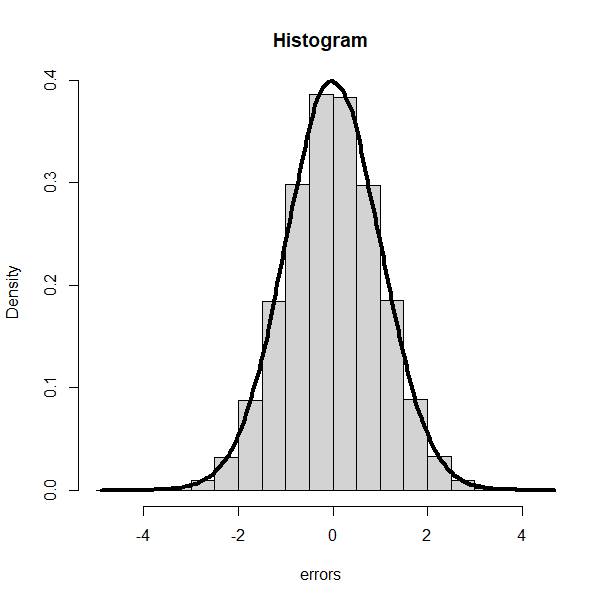} &
			\includegraphics[width=0.45\linewidth,height=3.5cm]{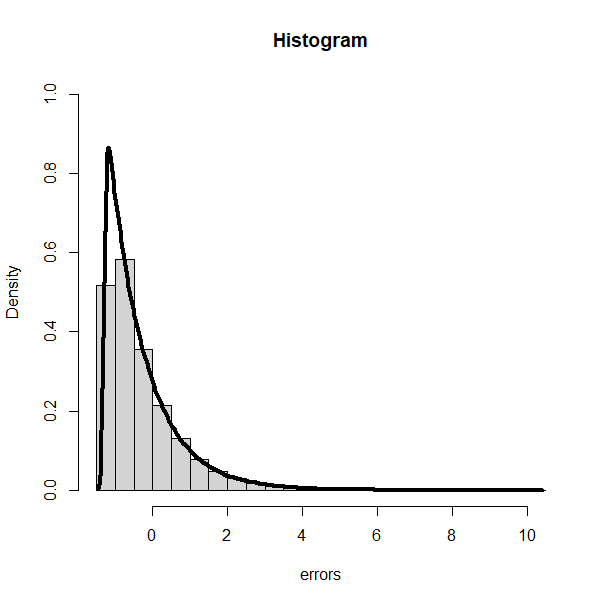} \\
				{\small (a) ${\cal N}(0,1)$.} &
			{\small (b) ${\cal E}xp(1)-1.3$.}\\
		& \\
		\includegraphics[width=0.45\linewidth,height=3.5cm]{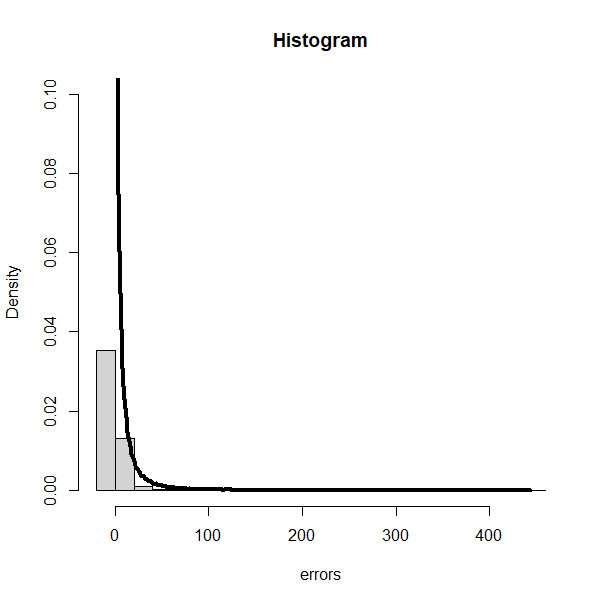} &
		\includegraphics[width=0.45\linewidth,height=3.5cm]{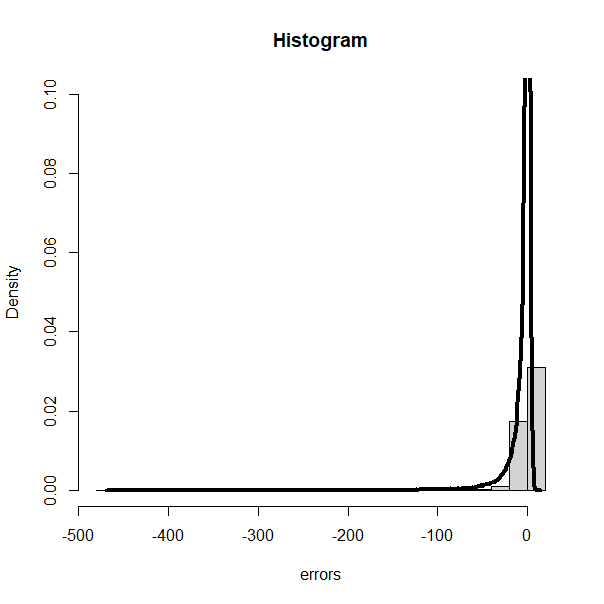} \\
			{\small (c) $({\cal N}(0,1))^4 - 6\,\text{median}(({\cal N}(0,1))^4)$.} &
			{\small (d) ${\cal E}xp(1) - ({\cal N}(0,1))^4$.} 
	\end{tabular}
		\caption{\small Histograms and density estimations for four distributions of $\varepsilon$.}
	\label{hists} 
\end{figure}

In this section, a numerical study is conducted to illustrate the theoretical results and to compare the model estimators   with those derived from the LS and quantile frameworks. For simulation study we use the following R language packages: package {\it SALES} for expectile and LS frameworks, package {\it quantreg} for the quantile.\\
Considered model is the following:
\begin{equation}
	Y_i =\sum^p_{j=1}\beta^*_j X_{ji} +\varepsilon_i, \qquad i=1, \cdots , n.
	\label{sm}
\end{equation}
For realizations of a given distribution of $\varepsilon_i$, we can calculate    the following empirical estimator  for the expectile index  $\tau$, taking into account assumption (A1):
\begin{equation*}
	\label{etae}
	\widehat \tau_{exp}=\frac{n^{-1}\sum^n_{i=1}\varepsilon_i  \e1_{\varepsilon_i <0}}{n^{-1} \big( \sum^n_{i=1} \varepsilon_i  \e1_{\varepsilon_i <0}-\sum^n_{i=1} \varepsilon_i  \e1_{\varepsilon_i   >0}\big)}.
\end{equation*}
Four model errors are considered: ${\cal N}(0,1)$, ${\cal E}xp(1)-1.3$, $({\cal N}(0,1))^4 - 6\,\text{median}(({\cal N}(0,1))^4)$, ${\cal E}xp(1) - ({\cal N}(0,1))^4$, for which the estimated expectile indices  $\hat{\tau}_{exp}$    are  $ 0.50, 0.678, 0.231, 0.806  $, respectively (see Figure \ref{hists} for their histograms and density estimations). The quantile index is fixed at $0.5$ and so this is a median regression. \\
The $n$ observations of $(Y_i,\eX_i)_{1 \leqslant i \leqslant n}$ are randomly split into training and test sets based on fixed percentages.\\
In order to calculate the model parameters for the estimation of $\widehat {\cal M}_n$ that minimises the CVS, the following methods are employed:  expectile, quantile or LS method. For estimators $\widehat{\cal M}^{(a)}_n$ which minimize aCVS, the model parameters are calculated by adaptive LASSO expectile, quantile or LS estimators. For each Monte Carlo replication we calculate the following three indicators: 
\begin{itemize}
	\item Mean squared error (MSE) on the validation set:
	\[
	\text{MSE}_{\cal V} = \frac{1}{n_{\cal V}}\sum_{i\in \mathcal{V}}(Y_i - \hat Y_i)^2,
	\]
\end{itemize}
In order to identify both relevant and irrelevant explanatory variables, two indicators are calculated for  $\widehat{\cal M}_n$ and $\widehat{\cal M}^{(a)}_n$:
\begin{itemize}
	\item True Positive Rate ($\text{TPR}_{\cal V}$): proportion of relevant variables correctly identified,
	\item True Negative Rate ($\text{TNR}_{\cal V}$): proportion of irrelevant variables correctly identified.
\end{itemize}
Let us emphasize that a perfect model has $\text{TPR}_{\cal V}$=$\text{TNR}_{\cal V}$=1.\\
The model is repeated for $N$ Monte Carlo replications, with the realizations of the random variables $(X_{ji})_{j \in \{1, \cdots, p\}}$ and $\varepsilon_i$ for $i=1, \cdots , n$ being repeated.  The indicators MSE, TPR and TNR  given in the tables are the arithmetic mean of the $MSE_{\cal V}$, $\text{TPR}_{\cal V}$ and $\text{TNR}_{\cal V}$ obtained for each Monte Carlo replication. \\
In Table \ref{tab_Fixed p} three values for $p$ are considered 2, 4, and 7, with corresponding true values  $\eb^*$: (2,0), (2,0,-1.5,-2), and (2,0,-1.5,-2,0,0,0), respectively. In order to examine the impact of assumption (A4), three values for the rate,  $s=n_{\cal V}/n$, are considered: 0.2, 0.8, and 0.9. A thorough examination of the results from the simulations for 0.8 and, more specifically, for 0.9 is warranted, as the theoretical underpinnings are substantiated when $n_{\cal V}/n_{\cal A}$ converges to infinity. Consequently, the rate $s$ approaches 1.
From the results presented in Table \ref{tab_Fixed p}, we can deduce that when $\varepsilon_i \sim {\cal N}(0, 1)$, the expectile and LS methods give the same results, with TPR and TNR increasing with the rate $s$.
When  $\varepsilon_i \sim {\cal E}xp(1)-1.3$, expectile performs less well than LS (when comparing TPR and TNR).
When distributions  $({\cal N}(0,1))^4 - 6\,\text{median}(({\cal N}(0,1))^4)$, ${\cal E}xp(1) - ({\cal N}(0,1))^4$, the expectile method gives better results for MSE, TPR, and TNR than the LS method. This can be interpreted as meaning that the model error distribution must be very asymmetric for that the model estimation by expectile framework to be  better  than LS. On the other hand, the obtained results for the quantile method for $s=0.2$ and $s=0.8$ are consistent with those of  \cite{Kim-Lee-Shin-Jung.25}. 
When $s=0.9$, the TNR for the quantile method converges more slowly toward 1. Therefore, for the same distribution of the model errors, the expectile method is more accurate for a given $n$.

\begin{table}
	\centering
	\renewcommand{\arraystretch}{1}
	\setlength{\tabcolsep}{5pt}
	\resizebox{\textwidth}{!}{
		\begin{tabular}{l| l  l ccc @{\hskip 20pt} ccc @{\hskip 20pt} ccc}
			\hline
			&    &  &  \multicolumn{3}{c}{\textbf{MSE}} 
			& \multicolumn{3}{c}{\textbf{TPR}} 
			& \multicolumn{3}{c}{\textbf{TNR}} \\
			\cline{4-6} \cline{7-9} \cline{10-12}
			$p$   & error & Method 
			& 20\% & 80\% & 90\% & 20\% & 80\% & 90\% & 20\% & 80\% & 90\% \\ 	\hline
			& ${\cal N}(0,1)$ & Expectile 	& 0.0020 & 0.0060 & 0.0106 	& 1.000 & 1.000 & 1.000 & 0.555 & 0.750 & 0.790 \\[2pt]			
			&  & LS & 0.0020 & 0.0060 & 0.0106 & 1.000 & 1.000 & 1.000 & 0.565 & 0.745 & 0.775 \\[2pt]
			&  & Quantile  & 0.0027 & 0.0091 & 0.0159 & 1.000 & 1.000 & 1.000 & 0.575 & 0.700 & 0.745 \\[3pt] \cdashline{2-12}\\[-7pt]
			&   ${\cal E}xp(1)-1.3$ & Expectile & 0.0023 & 0.0096 & 0.0162 & 1.000 & 1.000 & 1.000 & 0.590 & 0.685 & 0.775 \\[2pt]
			&   & LS & 0.0014 & 0.0055 & 0.0100 & 1.000 & 1.000 & 1.000 & 0.565 & 0.730 & 0.800 \\[2pt]
			&    & Quantile  & 0.0016 & 0.0056 & 0.0096 & 1.000 & 1.000 & 1.000 & 0.595 & 0.655 & 0.720 \\[3pt]
			\cdashline{2-12}\\[-7pt]
			2  & $({\cal N}(0,1))^4 - 6\,\mathrm{med}({\cal N}(0,1)^4)$ & Expectile 	& 0.0852 & 0.0765 & 0.2124 	& 0.970 & 1.000 & 0.980 & 0.650 & 0.705 & 0.755 \\[2pt]
			&   & LS  & 0.4408 & 0.4128 & 1.0205 & 0.860 & 0.940 & 0.875 & 0.580 & 0.645 & 0.695 \\[2pt]
			&   & Quantile  & 0.0016 & 0.0073 & 0.0157 & 1.000 & 1.000 & 1.000 & 0.635 & 0.675 & 0.680 \\[3pt]
			\cdashline{2-12}\\[-7pt]
			&  ${\cal E}xp(1)–({\cal N}(0,1))^4$ & Expectile & 0.0852 & 0.0765 & 0.2124 & 0.970 & 1.000 & 0.980 & 0.650 & 0.705 & 0.755 \\[2pt]
			&   & LS & 0.4408 & 0.4128 & 1.0205 & 0.860 & 0.940 & 0.875 & 0.580 & 0.645 & 0.695 \\[2pt]
			&    & Quantile  & 0.0016 & 0.0073 & 0.0157 & 1.000 & 1.000 & 1.000 & 0.635 & 0.675 & 0.680 \\[3pt]
			\hline 	\hline
			&  ${\cal N}(0,1)$ & Expectile 	& 0.0021 & 0.0079 & 0.0155 & 1.000 & 1.000 & 1.000 	& 0.610 & 0.665 & 0.790 \\[2pt]
			&  & LS & 0.0021 & 0.0078 & 0.0154 & 1.000 & 1.000 & 1.000 & 0.615 & 0.645 & 0.790 \\[2pt]
			&  & Quantile 	& 0.0031 & 0.0124 & 0.0228 	& 1.000 & 1.000 & 1.000 & 0.590 & 0.650 & 0.655 \\[3pt]
			\cdashline{2-12}\\[-7pt]
			&   ${\cal E}xp(1)-1.3$ & Expectile & 0.0037 & 0.0135 & 0.0272 & 1.000 & 1.000 & 1.000 	& 0.610 & 0.715 & 0.735 \\[2pt]
			&  & LS & 0.0021 & 0.0079 & 0.0169 & 1.000 & 1.000 & 1.000 & 0.615 & 0.740 & 0.740 \\[2pt]
			&  & Quantile 	& 0.0020 & 0.0070 & 0.0144 	& 1.000 & 1.000 & 1.000 & 0.520 & 0.695 & 0.640 \\[3pt]
			\cdashline{2-12}\\[-7pt]
			4   & $({\cal N}(0,1))^4 - 6\,\mathrm{med}({\cal N}(0,1)^4)$ & Expectile 	& 0.0902 & 0.1436 & 0.2626 	& 0.973 & 0.997 & 0.960  & 0.560 & 0.780 & 0.685 \\[2pt]
			&    & LS & 0.4992 & 0.6757 & 0.8374 & 0.865 & 0.867 & 0.825 & 0.570 & 0.750 & 0.720 \\[2pt]
			&  & Quantile  & 0.0015 & 0.0061 & 0.0161 & 1.000 & 1.000 & 1.000 & 0.515 & 0.645 & 0.700 \\[3pt]
			\cdashline{2-12}\\[-7pt]
			&   ${\cal E}xp(1) – ({\cal N}(0,1))^4$ & Expectile & 0.0851 & 0.1348 & 0.2619 & 0.977 & 0.998 & 0.967 & 0.525 & 0.745 & 0.750 \\[2pt]
			&   & LS & 0.5098 & 0.6670 & 0.8919 & 0.857 & 0.880 & 0.803 & 0.560 & 0.720 & 0.765 \\[2pt]
			&  & Quantile 	& 0.0023 & 0.0136 & 0.0348 & 1.000 & 1.000 & 1.000 	& 0.595 & 0.695 & 0.740 \\[3pt]
			\hline \hline
			&  ${\cal N}(0,1)$ & Expectile & 0.0016 & 0.0051 & 0.0107 & 1.000 & 1.000 & 1.000 & 0.599 & 0.721 & 0.782 \\[2pt]
			&   & LS & 0.0016 & 0.0051 & 0.0107 & 1.000 & 1.000 & 1.000 & 0.600 & 0.730 & 0.781 \\[2pt]
			&   & Quantile 	& 0.0024 & 0.0073 & 0.0140 	& 1.000 & 1.000 & 1.000 & 0.573 & 0.699 & 0.690 \\[3pt]
			\cdashline{2-12}\\[-7pt]
			&   ${\cal E}xp(1)-1.3$ & Expectile & 0.0029 & 0.0089 & 0.0150 & 1.000 & 1.000 & 1.000 & 0.573 & 0.725 & 0.784 \\[2pt]
			&  & LS & 0.0017 & 0.0052 & 0.0095 & 1.000 & 1.000 & 1.000 & 0.584 & 0.739 & 0.787 \\[2pt]
			&   & Quantile 	& 0.0016 & 0.0041 & 0.0084 	& 1.000 & 1.000 & 1.000 & 0.547 & 0.664 & 0.693 \\[3pt]
			\cdashline{2-12}\\[-7pt]
			7 &   $({\cal N}(0,1))^4 - 6\,\mathrm{med}({\cal N}(0,1)^4)$ & Expectile 	& 0.0481 & 0.0950 & 0.1982 	& 0.980 & 0.988 & 0.957 	& 0.630 & 0.741 & 0.795 \\[2pt]
			&   & LS & 0.3030 & 0.4434 & 0.6122 & 0.867 & 0.863 & 0.797 & 0.613 & 0.726 & 0.780 \\[2pt]
			&   & Quantile & 0.0011 & 0.0038 & 0.0078 & 1.000 & 1.000 & 1.000 & 0.591 & 0.650 & 0.689 \\[3pt]
			\cdashline{2-12}\\[-7pt]
			&   ${\cal E}xp(1) – ({\cal N}(0,1))^4$ & Expectile & 0.0667 & 0.0896 & 0.1716 & 0.967 & 0.992 & 0.962 & 0.608 & 0.735 & 0.789 \\[2pt]
			&   & LS & 0.3531 & 0.4350 & 0.5453 & 0.840 & 0.878 & 0.813 & 0.601 & 0.709 & 0.772 \\[2pt]
			&   & Quantile 	& 0.0014 & 0.0064 & 0.0175 	& 1.000 & 1.000 & 1.000 & 0.581 & 0.694 & 0.716 \\[3pt]
			\hline
		\end{tabular}
	}
	\caption{Results obtained  for MSE, TPR, and TNR by $\widehat{\cal M}_n$ of (\ref{eq3}), for different dimensions of $\eb^*$: 
		$p=2$ with $\eb^*=(2, 0)$, $p=4$ with $\eb^*=(2, 0, -1.5, -2)$, and $p=7$ with $\eb^*=(2, 0, -1.5, -2, 0, 0, 0)$. 
		All results are based on $N=200$ Monte Carlo replications and $n=500$.}
	\label{tab_Fixed p}
\end{table}

\begin{table}
	\centering
	\renewcommand{\arraystretch}{0.9}\setlength{\tabcolsep}{5pt}
	\resizebox{\textwidth}{!}{
		\begin{tabular}{ l l l cc @{\hskip 20pt} cc @{\hskip 20pt} cc}
			\hline
			$n$ & error & Method & \multicolumn{2}{c}{\textbf{MSE}} & \multicolumn{2}{c}{\textbf{TPR}} & \multicolumn{2}{c}{\textbf{TNR}} \\
			\cline{4-5} \cline{6-7} \cline{8-9}
			& & & 80\% & 90\% & 80\% & 90\% & 80\% & 90\% \\
			\hline
			& & Expectile & 0.0508 & 0.144 & 1.000 & 0.995 & 0.735 & 0.715 \\[2pt]
			& $\mathcal{N}(0,1)$ & LS & 0.0500 & 0.1421 & 1.000 & 0.995 & 0.725 & 0.735 \\[2pt]
			&  & Quantile & 0.0670 & 0.1912 & 1.000 & 0.992 & 0.685 & 0.700 \\
			\cdashline{2-9}\\[-7pt]
			& & Expectile & 0.0750 & 0.1802 & 0.998 & 0.987 & 0.725 & 0.745 \\[2pt]
			& ${\cal E}xp(1)-1.3$ & LS & 0.0538 & 0.1473 & 0.998 & 0.992 & 0.725 & 0.750 \\[2pt]
			&  & Quantile & 0.0496 & 0.1353 & 1.000 & 0.995 & 0.695 & 0.675 \\
			\cdashline{2-9}\\[-7pt]
		100	& & Expectile & 0.7952 & 0.9239 & 0.852 & 0.805 & 0.760 & 0.735 \\[2pt]
			& $(\mathcal{N}(0,1))^4 - 6\,\mathrm{med}(\mathcal{N}(0,1)^4)$ & LS & 1.6470 & 1.7049 & 0.685 & 0.642 & 0.710 & 0.665 \\[2pt]
			&  & Quantile & 0.1204 & 0.4463 & 0.982 & 0.918 & 0.740 & 0.730 \\
			\cdashline{2-9}\\[-7pt]
			& & Expectile & 0.7454 & 1.0503 & 0.837 & 0.753 & 0.730 & 0.760 \\[2pt]
			& ${\cal E}xp(1) - (\mathcal{N}(0,1))^4$ & LS & 1.6139 & 1.9022 & 0.623 & 0.645 & 0.660 & 0.710 \\[2pt]
			&  & Quantile & 0.2377 & 0.6993 & 0.972 & 0.860 & 0.695 & 0.780 \\
			\hline 
			& & Expectile & 0.0078 & 0.0155 & 1.000 & 1.000 & 0.693 & 0.790 \\[2pt]
			& $\mathcal{N}(0,1)$ & LS & 0.0078 & 0.0154 & 1.000 & 1.000 & 0.698 & 0.790 \\[2pt]
			&  & Quantile & 0.0101 & 0.0228 & 1.000 & 1.000 & 0.660 & 0.655 \\
			\cdashline{2-9}\\[-7pt]
			&   & Expectile & 0.0108 & 0.0272 & 1.000 & 1.000 & 0.762 & 0.735 \\[2pt]
			&  ${\cal E}xp(1)-1.3$& LS & 0.0065 & 0.0169 & 1.000 & 1.000 & 0.785 & 0.740 \\[2pt]
			&  & Quantile & 0.0060 & 0.0144 & 1.000 & 1.000 & 0.652 & 0.640 \\
			\cdashline{2-9}\\[-7pt]
			500 & & Expectile & 0.1094 & 0.2626 & 0.993 & 0.960 & 0.748 & 0.685 \\[2pt]
			& $(\mathcal{N}(0,1))^4 - 6\,\mathrm{med}(\mathcal{N}(0,1)^4)$& LS        & 0.4931 & 0.8374 & 0.883 & 0.825 & 0.725 & 0.720 \\[2pt]
			&  & Quantile  & 0.0047 & 0.0161 & 1.000 & 1.000 & 0.650 & 0.700 \\
			\cdashline{2-9}\\[-7pt]
			&  & Expectile & 0.1095 & 0.2619 & 0.990 & 0.967 & 0.750 & 0.750 \\[2pt]
			& ${\cal E}xp(1) - (\mathcal{N}(0,1))^4$ & LS 
			& 0.4794 & 0.8919 & 0.897 & 0.803 & 0.735 & 0.765 \\[2pt]
			&  & Quantile 
			& 0.0085 & 0.0348 & 1.000 & 1.000 & 0.693 & 0.740 \\
			\hline
			& & Expectile & 0.0024 & 0.0053 & 1.000 & 1.000 & 0.749 & 0.792 \\[2pt]
			&  $\mathcal{N}(0,1)$  & LS        & 0.0024 & 0.0053 & 1.000 & 1.000 & 0.748 & 0.792 \\[2pt]
			&  & Quantile  & 0.0031 & 0.0073 & 1.000 & 1.000 & 0.706 & 0.653 \\
			\cdashline{2-9}\\[-7pt]
			& & Expectile & 0.0042 & 0.0090 & 1.000 & 1.000 & 0.757 & 0.767 \\[2pt]
			& ${\cal E}xp(1)-1.3$ & LS        & 0.0024 & 0.0054 & 1.000 & 1.000 & 0.744 & 0.788 \\[2pt]
			&  & Quantile  & 0.0020 & 0.0047 & 1.000 & 1.000 & 0.655 & 0.660 \\
			\cdashline{2-9}\\[-7pt]
		1000	& & Expectile & 0.0351 & 0.0961 & 1.000 & 0.990 & 0.767 & 0.782 \\[2pt]
			& $(\mathcal{N}(0,1))^4 - 6\,\mathrm{med}(\mathcal{N}(0,1)^4)$& LS        & 0.2181 & 0.3841 & 0.957 & 0.910 & 0.756 & 0.760 \\[2pt]
			&  & Quantile  & 0.0016 & 0.0039 & 1.000 & 1.000 & 0.669 & 0.690 \\
			\cdashline{2-9}\\[-7pt]
			& & Expectile & 0.0348 & 0.0975 & 1.000 & 0.995 & 0.764 & 0.787 \\[2pt]
			& ${\cal E}xp(1) - (\mathcal{N}(0,1))^4$ & LS        & 0.2199 & 0.4228 & 0.962 & 0.892 & 0.729 & 0.778 \\[2pt]
			&  & Quantile  & 0.0026 & 0.0085 & 1.000 & 1.000 & 0.690 & 0.718 \\
			\hline
		\end{tabular}
	}
	\caption{ Results obtained  for MSE, TPR, and TNR by $\widehat{\cal M}_n$ of (\ref{eq3}) when   dimension $p$ varies with  $n$, $p=O(n_{\cal A}^c)$
		for $c=1/3$, and $s \in \{0.8, 0.9\}$.  Values for $n$ and $p$: when $n=100$ we consider $p=4$ (for s=0.8) and $p=4$ (for s=0.9); when $n=500$ we consider $p=5$ (for s=0.8) and $p=4$ (for s=0.9);  when $n=1000$ we consider $p=7$ (for s=0.8) and $p=6$ (for s=0.9). }
	
	\label{NonFixedp_c13}
\end{table}

\begin{table}
	\centering
	\renewcommand{\arraystretch}{1}\setlength{\tabcolsep}{5pt}
	\resizebox{\textwidth}{!}{
		\begin{tabular}{ l l l cc @{\hskip 20pt} cc @{\hskip 20pt} cc}
			\hline
			$n$ & error & Method & \multicolumn{2}{c}{\textbf{MSE}} & \multicolumn{2}{c}{\textbf{TPR}} & \multicolumn{2}{c}{\textbf{TNR}} \\
			\cline{4-5} \cline{6-7} \cline{8-9}
			& & & 80\% & 90\% & 80\% & 90\% & 80\% & 90\% \\
			\hline
			& & Expectile & 0.0199 & 0.0600 & 1.000 & 0.997 & 0.727 & 0.770 \\[2pt]
			& $\mathcal{N}(0,1)$ & LS & 0.0196 & 0.0586 & 1.000 & 1.000 & 0.730 & 0.768 \\[2pt]
			&  & Quantile & 0.0260 & 0.0652 & 1.000 & 1.000 & 0.716 & 0.708 \\
			\cdashline{2-9}\\[-8pt]
			& & Expectile & 0.0311 & 0.0861 & 1.000 & 0.987 & 0.710 & 0.718 \\[2pt]
			&  ${\cal E}xp(1)-1.3$ & LS        & 0.0229 & 0.0714 & 1.000 & 0.990 & 0.734 & 0.752 \\[2pt]
			&  & Quantile  & 0.0152 & 0.0468 & 1.000 & 0.997 & 0.673 & 0.730 \\
			\cdashline{2-9}\\[-8pt]
			100 & & Expectile & 0.4046 & 0.4873 & 0.840 & 0.770 & 0.729 & 0.774 \\[2pt]
			& $(\mathcal{N}(0,1))^4 - 6\,\mathrm{med}(\mathcal{N}(0,1)^4)$ & LS        & 1.0411 & 0.8863 & 0.617 & 0.613 & 0.649 & 0.724 \\[2pt]
			& & Quantile  & 0.0424 & 0.2036 & 0.987 & 0.913 & 0.703 & 0.782 \\[3pt]
			\cdashline{2-9}\\[-8pt]
			& & Expectile & 0.3672 & 0.5840 & 0.840 & 0.733 & 0.741 & 0.784 \\[2pt]
			& ${\cal E}xp(1) - (\mathcal{N}(0,1))^4$  & LS        & 0.9102 & 1.1841 & 0.660 & 0.563 & 0.693 & 0.722 \\[2pt]
			&& Quantile  & 0.0984 & 0.2816 & 0.963 & 0.877 & 0.723 & 0.774 \\
			\hline 
			& & Expectile & 0.0067 & 0.0088 & 1.000 & 1.000 & 0.667 & 0.740 \\[2pt]
			& $\mathcal{N}(0,1)$ & LS        & 0.0067 & 0.0087 & 1.000 & 1.000 & 0.683 & 0.740 \\[2pt]
			&& Quantile  & 0.0079 & 0.0072 & 1.000 & 1.000 & 0.550 & 0.680 \\
			\cdashline{2-9}\\[-8pt]
			& &Expectile & 0.0042 & 0.012  & 1.000 & 1.000 & 0.800 & 0.720 \\[2pt]
			&${\cal E}xp(1)-1.3$& LS        & 0.0029 & 0.0073 & 1.000 & 1.000 & 0.767 & 0.720 \\[2pt]
			&& Quantile  & 0.0032 & 0.0041 & 1.000 & 1.000 & 0.617 & 0.800 \\
			\cdashline{2-9}\\[-8pt]
		500	& & Expectile & 0.0837 & 0.0510 & 1.000 & 1.000 & 0.800 & 0.760 \\[2pt]
			& $(\mathcal{N}(0,1))^4 - 6\,\mathrm{med}(\mathcal{N}(0,1)^4)$ & LS        & 0.3440 & 0.2155 & 0.800 & 0.867 & 0.717 & 0.760 \\[2pt]
			&& Quantile  & 0.0017 & 0.0025 & 1.000 & 1.000 & 0.650 & 0.720 \\
			\cdashline{2-9}\\[-8pt]
			&  & Expectile & 0.0393 & 0.0835 & 1.000 & 0.933 & 0.700 & 0.820 \\[2pt]
			& ${\cal E}xp(1) - (\mathcal{N}(0,1))^4$ & LS        & 0.1822 & 0.3265 & 0.933 & 0.733 & 0.717 & 0.700 \\[2pt]
			&& Quantile  & 0.0030 & 0.0061 & 1.000 & 1.000 & 0.767 & 0.640 \\
			\hline
			&$\mathcal{N}(0,1)$   & Expectile  & 0.0012 & 0.0048 & 1.000 & 1.000 & 0.80 & 0.820 \\[2pt]
			&& LS        & 0.0012 & 0.0048 & 1.000 & 1.000 & 0.80 & 0.820 \\[2pt]
			&& Quantile  & 0.0019 & 0.0061 & 1.000 & 1.000 & 0.710 & 0.794 \\
			\cdashline{2-9}\\[-8pt]
			& & Expectile & 0.0031 & 0.0048& 1.000 & 1.000 & 0.799 & 0.828 \\[2pt]
			&  ${\cal E}xp(1)-1.3$ & LS        & 0.0023& 0.0035 & 1.000 & 1.000 & 0.796 & 0.848 \\[2pt]
			&  & Quantile  & 0.0011 &  0.0019 & 1.000 & 1.000 & 0.755 & 0.801 \\
			\cdashline{2-9}\\[-8pt]
		1000	& & Expectile & 0.0234 &  0.0305  & 1.000 & 0.990 & 0.867 & 0.959 \\[2pt]
			& $(\mathcal{N}(0,1))^4 - 6\,\mathrm{median}(\mathcal{N}(0,1)^4)$& LS       & 0.0162  & 0.1787 & 1.000 & 1.000 & 0.802 & 0.839 \\[2pt]
			&  & Quantile  &0.0012  & 0.0021  & 1.000 & 1.000 & 0.799 & 0.803 \\
			\cdashline{2-9}\\[-8pt]
			& & Expectile & 0.0302  & 0.0436  & 1.000 & 1.000 & 0.988 &1.000 \\[2pt]
			& ${\cal E}xp(1) - (\mathcal{N}(0,1))^4$ & LS        & 0.1267 &  0.1396 & 1.000 & 1.000 &0.935 & 0.954 \\[2pt]
			&  & Quantile  & 0.0002 & 0.0009 & 1.000 & 1.000 & 0.870 & 0.889 \\
			\hline
		\end{tabular}
	}
	\caption{Results obtained  for MSE, TPR, and TNR by $\widehat{\cal M}_n$ of (\ref{eq3}) when   dimension $p$ varies with  $n$, $p=O(n_{\cal A}^c)$
		for  $c=4/5$, and $s \in \{0.8, 0.9\}$.  Values for $n$ and $p$: when $n=100$ we consider  $p=10$ (for s=0.8) and $p=8$ (for s=0.9);  when $n=500$ we consider  $p=20$ (for s=0.8) and $p=15$ (for s=0.9); when $n=1000$ we consider  $p=50$ (for s=0.8) and $p=30$ (for s=0.9).}
		\label{NonFixedP_p45}
\end{table}

In Tables \ref{NonFixedp_c13} and \ref{NonFixedP_p45},  the true coefficient vector $\eb^*$ is $(2, 0, -1.5, -2)$ for $p=4$ and for $p>4$ zeros are appended to match the dimension.   The results presented in Tables  \ref{NonFixedp_c13} and \ref{NonFixedP_p45} are obtained following 200 Monte Carlo replications for the same four distributions for $\varepsilon$ and quantile index equal to 0.5.  From these two tables, we can deduce that if $p$ depends on $n$, then the TNR increases with the number of zeros in the parameter vector. For $n=100$, the quantile (more precisely, the median) method  yields better results than the expectile method in terms of MSE and TPR but worse results in terms of TNR. For $n \in \{500, 1000\}$, the expectile and quantile results are similar with respect to TPR. However, with respect to TNR, the expectile method is better at detecting irrelevant variables.  As in the case of a fixed $p$, we find that the TNR obtained by the expectile method approaches 1 more quickly as the rate $s$ approaches 1. The identification of zero coefficients increases with the sample size. For asymmetric model errors, the LS method always identifies fewer irrelevant variables and does not identify all relevant ones, especially for small $n$. \\

Now, let us examine the model estimator for cross-validation mean score (\ref{ta}) calculated for the adaptive LASSO expectile estimator. Estimator (\ref{Man}) is compared with the corresponding to the adaptive LASSO median and LS estimators for the same four model errors following 200 Monte Carlo replications. The considered model in Table \ref{Fixed p_penalized case} has seven explanatory variables, three of which are relevant. \\
By comparing the results in Tables \ref{Fixed p_penalized case} and Table \ref{tab_Fixed p} for $p=7$, we see that the CV model estimators by the three techniques: expectile, quantile, LS, have almost identical TPRs for a given error distribution and technique. However, since  the adaptive LASSO satisfies the sparsity property, the  model $\widehat {\cal M}^{(a)}_n $ identifies irrelevant variables much better than the  model $\widehat {\cal M}_n $ for unpenalized estimators. In terms of accuracy, the MSE values are very close for both model estimators  $\widehat {\cal M}_n $ and $\widehat {\cal M}^{(a)}_n $.\\
In Table \ref{tab_comp_FS} we present the results when $p=10$ for two values of $\eb^*$: $\eb^0= (2,0,-1.5,0, 0,0,0,$ $0,0,0)$ and $\eb^1= (0.2,0,-0.5,0, 0,0,0,0,0,0)$ which has nonzero values closer to zero.  In this table, the results obtained by $\widehat {\cal M}^{(a)}_n $, denoted by {\it CV\_Exp, CV\_LS, CV\_Q},  are compared using the two criteria TPR and TNR when $n=100$ and $n_{\cal A}=15$. The number of learning observations is very close to the number of parameters. Moreover, the comparison is made with the adaptive LASSO estimators (denoted by {\it FS\_Exp, FS\_LS, FS\_Q}) for the three estimation methods on the all $n$ observations. The  CV estimators identify zero coefficients better and non-zero coefficients slightly less well than adaptive LASSO estimators on full size. It should be noted that  when non-zero  values of $\beta^*_j$ are close to zero, the CV expectile estimator identifies irrelevant variables very well,  unlike the CV LS, CV quantile, and adaptive LASSO methods.

\begin{table}
	\centering
	\renewcommand{\arraystretch}{1.2}
	\setlength{\tabcolsep}{5pt}
	\resizebox{\textwidth}{!}{
		\begin{tabular}{l l l l ccc @{\hskip 20pt} ccc @{\hskip 20pt} ccc}
			\hline
			&  &  &  &  \multicolumn{3}{c}{\textbf{MSE}} 
			& \multicolumn{3}{c}{\textbf{TPR}} 
			& \multicolumn{3}{c}{\textbf{TNR}} \\
			\cline{5-7} \cline{8-10} \cline{11-13}
			$p$ &  $n$ &  error & Method 	& 20\% & 80\% & 90\%& 20\% & 80\% & 90\%& 20\% & 80\% & 90\% \\	\hline
			& 500 & N(0,1)& Expectile & 0.0012 & 0.0051 & 0.0127& 1.000 & 1.000 & 1.000	& 1.000 & 1.000 & 1.000 \\[2pt]
			&  &  & LS	& 0.0012 & 0.0051 & 0.0126	& 1.000 & 1.000 & 1.000	& 1.000 & 1.000 & 1.000 \\[2pt]
			&  &  & Median & 0.0018 & 0.0081 & 0.0188	& 1.000 & 1.000 & 1.000 &1.000 & 0.993  & 0.985 \\
			\cdashline{1-13}\\[-7pt]
			& 500 & Exp(1)-1.3 & Expectile 	& 0.0022 & 0.0096 & 0.0180 	& 1.000 & 1.000 & 1.000 & 0.994 & 0.999 & 1.000 \\[2pt]
			&  &  & LS 	& 0.0013 & 0.0058 & 0.0116 	& 1.000 & 1.000 & 1.000 & 1.000 & 0.999 & 1.000 \\[2pt]
			&  &  & Median & 0.0011 &  0.0038  & 0.008 & 1.000 & 1.000 & 1.000 & 1.000 & 0.994 & 0.989 \\
			\cdashline{1-13}\\[-7pt]
			7 & 500 & $(N(0,1))^4 - 6\,\mathrm{med}(N(0,1)^4)$ & Expectile 	& 0.0450 & 0.0968 & 0.2085 	& 0.980 & 0.988 & 0.945
			& 0.887 & 0.946 & 0.964 \\[2pt]
			&  &  & LS 	& 0.2991 & 0.4497 & 0.6209 	& 0.863 & 0.843 & 0.768 & 0.772 & 0.858 & 0.886 \\[2pt]
			&  &  & Median & 0.0011 & 0.0043 & 0.009 & 1.000 & 1.000 & 1.000 & 1.000 & 0.989 & 0.991 \\
			\cdashline{1-13}\\[-7pt]
			& 500 & Exp(1) – $(N(0,1))^4$ & Expectile & 0.0631 & 0.0906 & 0.1852 & 0.967 & 0.992 & 0.945 & 0.876 & 0.929 & 0.963 \\[2pt]
			&  &  & LS & 0.3489 & 0.4452 & 0.5618 & 0.838 & 0.862 & 0.787 & 0.761 & 0.835 & 0.886 \\[2pt]
			&  &  & Median & 0.0012 & 0.0097 & 0.0372 & 1.000 & 1.000 & 1.000 & 1.000 & 0.994 & 0.971  \\
			\hline
		\end{tabular}
	}
	\caption{Results for MSE, TPR, and TNR when $p=7$, with $\eb^*=(2, 0, -1.5, -2, 0, 0, 0)$. 
		Results are based on $N=200$ Monte Carlo replications for the model estimator $\widehat {\cal M}^{(a)}_n $ of relation (\ref{Man}). }
	\label{Fixed p_penalized case}
\end{table}
\begin{table}
	\centering
	\begin{tabular}{l |c c| cc}
		\hline
	\textbf{Method}	& \multicolumn{2}{c}{\textbf{ Results when $\eb^{*}=\eb^0$}}& \multicolumn{2}{|c}{\textbf{ Results when $\eb^{*}=\eb^1$}}  \\
			\cdashline{2-5}
		 & TPR & TNR  & TPR & TNR \\
		\hline
		{\it CV\_Exp}  & 0.9 & 1.000  & 0.8 & 0.992 \\
		{\it  FS\_Exp}  & 1.0 & 0.901  & 0.88 & 0.900 \\
		{\it  CV\_LS}   & 0.8 & 0.900   & 0.76 & 0.875 \\
		{\it  FS\_LS}   & 0.9 & 0.765 & 0.9 & 0.702  \\
		{\it  CV\_Q}    & 1.0 & 0.950  & 0.87 & 0.910  \\
		{\it  FS\_Q}     & 1.0 & 1.000   & 0.87 & 0.991 \\		\hline
	\end{tabular}
	\caption{Results on   TPR and TNR, for CV estimators (denoted "{\it CV\_..."}) and adaptive LASSO estimators on all observations (denoted "{\it FS\_..."})). Results following 10 Monte Carlo replications, with $n=100$, $n_{\cal A}=15$, $n_{\cal V}=85$, $p=10$, $\eb^0= (2,0,-1.5,0, 0,0,0,0,0,0)$ and $\eb^1= (0.2,0,-0.5,0, 0,0,0,0,0,0)$.  }
	\label{tab_comp_FS}
\end{table}
   
\begin{table}
	\centering
	\begin{tabular}{c|c|c|c}
		\hline
		\textbf{Dimension $p$} & \textbf{Method} & \multicolumn{2}{c}{\textbf{Computation time (seconds / hours)}} \\
		\cline{3-4}
			 & & {\textbf{unpenalized}} &  {\textbf{penalized}} \\
		\hline
		\multirow{2}{*}{10}
		& $CV Quantile$   & 7.68  ($\approx$  0.002 h) &  8.05  ($\approx$  0.002 h)\\
		& $CV Expectile$  & 6.38  ($\approx$  0.002 h) &  6.38  ($\approx$  0.002 h) \\ 
		&  $full Expectile $  & 3.54 ($\approx$ 0.0008 h) \\
		\hline
		\multirow{2}{*}{18}
		& $CV Quantile$   & 1819 ($\approx$  0.505 h) &  1858  ($\approx$  0.516 h) \\
		& $CV Expectile$  & 1454  ($\approx$  0.403 h)  &  1977  ($\approx$  0.549 h)\\
		\hline
		\multirow{2}{*}{20}
		& $CV Quantile$   & 10740  ($\approx$  2.983 h) & 12776  ($\approx$  3.548 h)\\
		& $CV Expectile$  & 7288 ($\approx$  2.024 h)  &   11613 ($\approx$   3.225 h)\\
			& $full Expectile$  &  6488  ($\approx$ 1.802 h)  \\
		\hline
		\multirow{2}{*}{30}
		& $CV Quantile$   & 57000 ($\approx$  15.833 h) & 63201 ($\approx$   17.555 h)\\
		& $CV Expectile$  & 29650 ($\approx$  8.236 h)&   35120 ($\approx$   9.722 h)\\
		\hline
			\end{tabular}
		\caption{Computation times for the CV quantile and expectile methods for non-penalized and penalized estimators. Calculation time to estimate a model using the unpenalized expectile method on all  $n$ observations (full). We consider  $n=100$, one Monte Carlo replication, $n_{\cal A}=20$, different values of $p$.}
		\label{tab_comparison}
	\end{table}
Let us now consider the computation time required to perform a Monte Carlo loop for each of the methods, either by cross-validation or on the entire database, for penalized or unpenalized estimators. The model under consideration has $p \in \{ 10, 18, 20, 30\}$ explanatory variables, only three of which are relevant. From Table \ref{tab_comparison}, we can deduce that the compution time using the CV method is shorter per expectile than per quantile, whether for penalized or non-penalized estimators. This difference in computation time increases with the value of $p$. We can also deduce that, for expectile, the computation time over all observations is smaller than for cross-validation, but this difference decreases as $p$ increases. The results in this table reinforce the advantage of the CV expectile method  over the CV quantile method, especially for a model with a large number of explanatory variables.
\section{Application on real data}
\label{section_applications}
In order to support the practical interest of our model selection method,  in this section, we present two applications on real databases.  The results are compared to those obtained by   median  and least squares cross validation. \\
Before presenting the applications, note that the value of the expectile index, $\tau$, ois unknown in these cases since $\varepsilon_i$ are not observable. Because   $(Y_i, \eX_i)_{1 \leqslant i \leqslant n}$ are observable, with the observations denoted $(y_i, \ex_i)_{1 \leqslant i \leqslant n}$,  a method to estimate the value of the expectile index. \\
Based on assumption (A7) and the dependence of $\eX$ on $Y$, we propose  the following schema  to calculate the empirical estimation of $\tau$ based on  assumption (A1). \\
{\it Step 0.} Initially we consider  $\eta^{(0)}(y_i) \equiv y_i-median( y_1, \cdots, y_n) $ which is used to calculate: 	
\begin{equation}
	\label{eta0}
	\widehat \tau^{(0)}=\frac{n^{-1}\sum^n_{i=1}\eta^{(0)}(y_i) \e1_{\eta^{(0)}(y_i)<0}}{n^{-1} \big( \sum^n_{i=1} \eta^{(0)}(y_i) \e1_{\eta^{(0)}(y_i)<0}-\sum^n_{i=1} \eta^{(0)}(y_i) \e1_{\eta^{(0)}(y_i) >0}\big)}.
\end{equation}
{\it Step 1.} Using the index  $\widehat \tau^{(0)}$, the  expectile estimations of the model parameters are calculated from which we calculate the previsions $\widehat y_i^{(1)}$ of $y_i$ and the corresponding residuals $e_i^{(1)} \equiv y_i - \widehat y_i^{(1)}$.   We calculate: 
\begin{equation}
	\label{etak}
	\widehat \tau^{(1)}=\frac{n^{-1}\sum^n_{i=1}e_i^{(1)} \e1_{e_i^{(1)}<0}}{n^{-1} \big( \sum^n_{i=1} e_i^{(1)} \e1_{e_i^{(1)}<0}-\sum^n_{i=1} e_i^{(1)} \e1_{e_i^{(1)}  >0}\big)}.
\end{equation}
{\it Step 2.}  We will consider  $\widehat \tau =	\widehat \tau^{(1)}$ as the estimated value for $\tau$. \\

\subsection{Acute aquatic toxicity}
\begin{figure}[h!] 
 		\includegraphics[width=1.\linewidth,height=4.5cm]{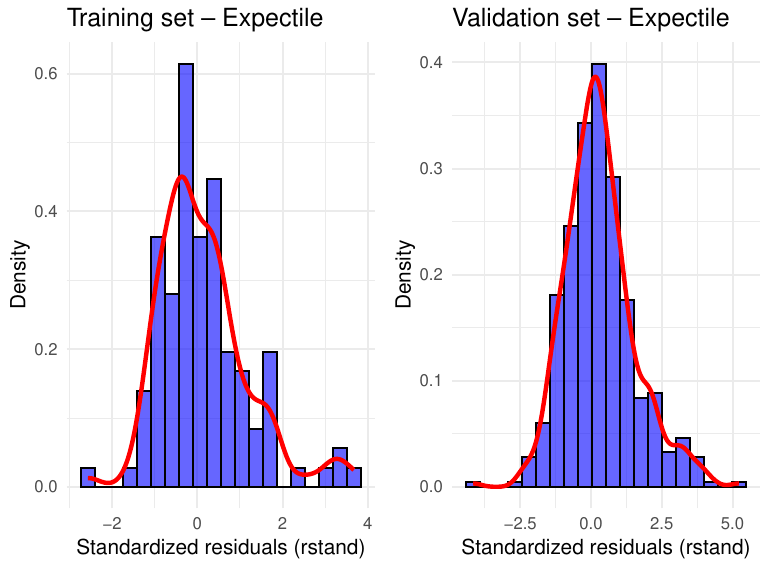 }
	\caption{Acute aquatic toxicity data: standardized residuals corresponding to the CV expectile method on the training and validation sets.	}
	\label{fig_aqua_hist} 
\end{figure}

\noindent We consider the following data:  \href{http://archive.ics.uci.edu/dataset/505/qsar+aquatic+toxicity}{\textit{http://archive.ics.uci.edu/dataset/505/qsar+aquatic+toxicity}}.\\
The  data set consists on  546 observations to predict acute aquatic toxicity toward Daphnia magna  denoted by \texttt{LC50}.  The eight continuous  explanatory  variables of molecular descriptors  are:  total polar surface area \texttt{TPSA(Tot)},  number of hydrogen bond acceptor atoms \texttt{SAacc},  fragments centered on an atom \texttt{H-050},  Moriguchi log P partition coefficient \texttt{MLOGP},  connectivity indices Connectivity indices \texttt{RDCHI},  2D autocorrelations \texttt{GATS1p},  constitutional indices \texttt{nN} and  atom-centred fragments \texttt{C-040}. 
A complete description of the data can be found in the paper of \cite{Cassotti.14}.  
The data were divided into a training set (20$\%$) and a validation set (80$\%$).  Variable selection was performed by cross-validation (\ref{eq3})  using three parameter estimation methods  on the training set: expectile, LS and median.  The results summarized in Table \ref{tab_aqua_estim} show that the expectile and LS models retained the variables \texttt{TPSA(Tot), SAacc, MLOGP, RDCHI, GATS1p, nN}, while the quantile model selected \texttt{TPSA(Tot), SAacc, RDCHI, GATS1p, nN}.    The estimation of the expectile index  $\tau$ calculated by (\ref{etak}) is 0.41, indicating asymmetry of residuals.
\begin{table}
	\centering
		\renewcommand{\arraystretch}{1}\setlength{\tabcolsep}{4pt}
		\begin{tabular}{lcccccccc} 		\hline
		Method & TPSA(Tot) & SAacc & H-050 & MLOGP & RDCHI & GATS1p & nN & C-040 \\
		\hline
		Expectile & 0.0277 & -0.0141 & 0 & 0.5301 & 0.3848 & -0.5353 & -0.2107 & 0 \\
		Least Squares & 0.0282 & -0.0144 & 0 & 0.5268 & 0.3873 & -0.5235 & -0.1917 & 0 \\
		Quantile & 0.0283 & -0.0108 & 0 & 0.7077 & 0 & -0.5631 & -0.1718 & 0 \\
		\hline
	\end{tabular}
	\caption{Acute aquatic toxicity data, selected variables and estimated coefficients  by: CV expectile, CV LS and CV median methods.}
	\label{tab_aqua_estim}
\end{table}
Note that the residuals for the models selected by the three estimation methods are not normally distributed. The Shapiro-Wilk test p-value for each method is less than $10^{-4}$ on the training set and less than $10^{-7}$ on the validation set (see Figure \ref{fig_aqua_hist} for the CV expectile method).

Now, let us compare the median and LS models selected  by cross-validation with the two  corresponding models built on the entire database of $n$ observations. The results obtained by these latter models are shown in Table \ref{tab_aqua_coefficients}.  Note that in the R SALES package does not offer any hypothesis tests on the expected estimates of the coefficients.   Hence another important aspect of the method proposed in this paper is that it allows estimates of both non-zero and zero coefficients to be obtained directly.  Comparing Tables \ref{tab_aqua_estim} and \ref{tab_aqua_coefficients}, we observe that the variable \texttt{RDCHI} was not retained by the median cross-validation method, even though it is significant across the entire data set. The CV expectile  method correctly identifies the significant variables, while the CV LS  method cannot be used because the residuals are not Gaussian.

\begin{table}[H]
	\centering
	\large
	\resizebox{\textwidth}{!}{%
		\begin{tabular}{l|cccc|cccc} 			\hline
			\textbf{Explanatory} & \multicolumn{4}{c|}{\textbf{LS}} &   \multicolumn{4}{c}{\textbf{Median}} \\
			\cline{2-9}
			\textbf{variables} 	& Estimate & Std. Err & t value & Pr($>|t|$)  & Estimate & Std. Err & Z value & p-value \\ 		\hline
		Intercept & 2.70 & 0.24 & 11.01 & $< 2\times10^{-16}$ 		& 2.84 & 0.18 & 15.54 & 0.000 \\ 		
		TPSA\_Tot & 0.02 & 0.0027 & 10.21 & $< 2\times10^{-16}$ 		& 0.02 & 0.002 & 14.4 & 0.000 \\
			SAacc & -0.015 & 0.002 & -7.19 & $2\times10^{-12}$ 	& -0.016 & 0.001 & -10.72 & 0.000 \\
				H\_050 & 0.04 & 0.06 & 0.67 & 0.50
		& 0.087 & 0.04 & 1.95 & 0.05 \\
			MLOGP & 0.44 & 0.06 & 7.04 & $5\times10^{-12}$ 		& 0.45 & 0.047 & 9.71 & 0.000 \\
				RDCHI & 0.51 & 0.13 & 3.770 & 0.0001 		& 0.47 & 0.10 & 4.71 & 0.000 \\
				GATS1p & -0.57 & 0.15 & -3.7 & 0.0002 		& -0.71 & 0.11 & -6.23 & 0.000 \\
				nN & -0.22 & 0.048& -4.65 & $4\times10^{-6}$ 	& -0.27 & 0.036 & -7.56 & 0.000 \\
				C\_040 & 0.003 & 0.078 & 0.043 & 0.96 	& -0.027 & 0.058 & -0.468 & 0.64 \\ 	\hline
		\end{tabular}
	}
	\caption{Acute aquatic toxicity data: estimations,  standard errors, test statistic value and p-values for  LS and median regressions on full data. The LS and median residuals are not normal. (Shapiro-Wilk p-values are $1.06\times10^{-8}$ et $3.39\times10^{-10}$, respectively).}
	\label{tab_aqua_coefficients}
\end{table}

\subsection{Energy Consumption}
\noindent We consider the following data: \\ \href{https://archive.ics.uci.edu/dataset/851/steel+industry+energy+consumption}{\textit{https://archive.ics.uci.edu/dataset/851/steel+industry+energy+consumption}}.\\
The  data set consists on  35040 observations to predict Steel Industry Energy Consumption  denoted by \texttt{Usage\_kWh}.  The six continuous  explanatory  variables describing production conditions, the temporal context, and characteristics of the industrial environment are: reactive energy lag \texttt{Lagging\_kVarh},  anticipated reactive energy \texttt{Leading\_kVarh},  carbon dioxide emissions associated with production \texttt{CO2},  power factor in delayed mode \texttt{Lagging\_Factor},  power factor in advance mode \texttt{Leading\_Factor} and   number of seconds from midnight \texttt{NSM}.  The estimation $\widehat \tau$ of the expectile index is 0.11 which shows that the model errors are highly asymmetrical.

\begin{table}[H]
	\centering
	\resizebox{\textwidth}{!}{%
		\begin{tabular}{lcccccc}
			\hline
			Method & Lagging\_kVarh & Leading\_kVarh & CO2 & Lagging\_Factor & Leading\_Factor & NSM \\
			\hline
			CV Expectile & 0.1613 & -0.206 & 1749.074 & 0 & -0.0678 & 0 \\
			CV LS & 0.4001 & 0.1675 & 1560.351 & 0.1656 & 0.0992 & 0 \\
			CV median & 0.279 & 0.1769 & 1748.261 & 0.0655 & 0.0535 & 0 \\
			\hline
		\end{tabular}%
	}
	\caption{Energy Consumption data: selected variables and estimated coefficients by: CV expectile, CV LS and CV median methods.}
\label{tab_energy_estim}
\end{table}

\begin{table}[H]
	\centering
	\large
	\resizebox{\textwidth}{!}{
		\begin{tabular}{l|c|cccc|ccccc} 	\hline
			 \multicolumn{1}{c|}{\textbf{Explanatory}} &  \multicolumn{1}{c|}{\textbf{Expectile}} &\multicolumn{4}{c|}{\textbf{LS}} &   \multicolumn{4}{c}{\textbf{Median}} \\
			\cline{2-10}
			\textbf{variables} & Estimate 	& Estimate & Std. Err & t value & Pr($>|t|$)  & Estimate & Std. Error & Z value & p-value  \\ 		\hline
			Intercept &12.57 &  -13.46 & 0.36 & -37.40 & $< 2\times10^{-16}$ & -7.16 & 0.052 & -136.04 & 0.000 \\
			Lagging\_kVarh &0.141 &  0.30 & 0.0038 & 78.050 & $< 2\times10^{-16}$ & 0.277 & 0.00057 & 486.609 & 0.000 \\
			Leading\_kVarh & -0.263  &  0.13 & 0.01 & 13 & $< 2\times10^{-16}$ 	& 0.17 & 0.001 & 113.91 & 0.000 \\
			CO2 & 1768.23& 1685.0 &  4.86 & 346.53 & $< 2\times10^{-16}$ & 1757.28 & 0.71 & 2469.54 & 0.000 \\
			Lagging\_Factor &  -0.004 & 0.12 & 0.0025 & 49.39 & $< 2\times10^{-16}$
			& 0.065 & 0.00038 & 175.357 & 0.000 \\
			Leading\_Factor & -0.075   & 0.075 & 0.0027 & 28.03 & $< 2\times10^{-16}$
			& 0.053 & 0.0004 & 136.153 & 0.000 \\
			NSM &  $7 \times 10^{-10}$ & 0.000009 & 0.000001 & 7.42 & 1.17$\times 10^{-13}$
			& 0.0000018 & 0.00000018& 9.82 & $< 2\times10^{-16}$ \\
			\hline
						\end{tabular}
	}
	\caption{Energy Consumption data: estimations, standard errors, t-values and p-values for expectile, LS and median methods on full data.}
	\label{tab:coefficients_steel}
\end{table}

\begin{figure}[h!] 
	\includegraphics[width=1.\linewidth,height=4.5cm]{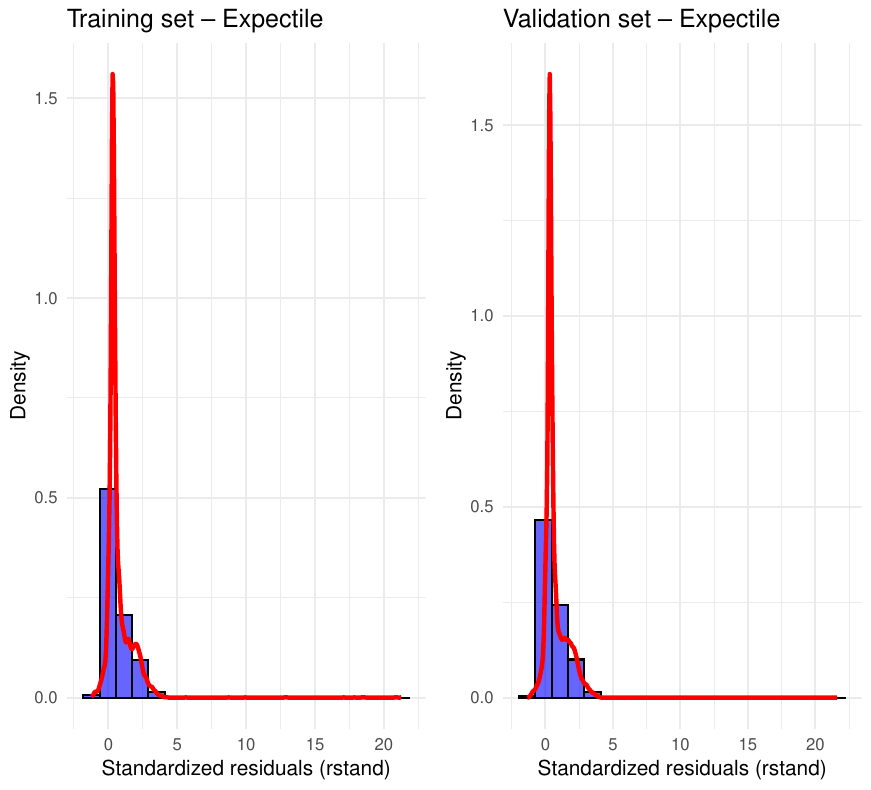 }
	\caption{Energy Consumption data: standardized residuals corresponding to the CV expectile method on the training and validation sets.	}
	\label{fig_energy_hist} 
\end{figure}

The standardized residuals on the training and validation sets using the CV expectile method are not normally distributed (see histograms of Figure \ref{fig_energy_hist}). The variable NSM is not selected by the, CV exoectile,  CV LS and CV median methods, but it is given as significant by the LS and median methods on the entire data set (see Tables \ref{tab_energy_estim} and \ref{tab:coefficients_steel}). This can be explained by the fact that the residuals obtained by the LS method do not follow a normal distribution and by the estimation very close to 0 of the corresponding coefficient obtained using the median method.\\
Regarding the opposite signs of the coefficient estimations obtained using the Regarding the opposite signs of the coefficient estimates obtained using the expectile CV method compared to those obtained using the median CV and LS CV, we offer the following explanation.CV Regarding the opposite signs of the coefficient estimates obtained using the expectile CV method compared to those obtained using the median CV and LS CV, we offer the following explanation. method compared to those obtained using the  CV median and  CV LS, we offer the following explanation. First, CV LS  estimates are unreliable because the corresponding residuals are not normally distributed. Estimates using the CV quantile method correspond to estimation for median regression, which in the case of highly skewed errors may differ from those using the CV expectile.

\section{Conclusion and  CV expectile method strengths}
\label{section_conclusion}
In this paper, we propose and study theoretically and numerically the expectile approach using the Train-Test split method, a subject that, to the authors' knowledge,  has not before been treated in the literature. Moreover, unlike usual,  the training and validation sets have the property that the validation set is much larger than the training set. The expectile or adaptive expectile LASSO estimators of the model parameters are calculated using a randomly selected training set from the entire database. Afterwards, the optimal model is selected by minimizing the cross-validation mean score on the validation set. We demonstrated the model consistency in three scenarios based on the number of explanatory variables, which may or may not be related to the number of observations. The consistency of the CV expectile  model estimator means that we can directly identify the zero coefficients of the true model, and estimate those that are not zero, with a probability that converges to 1 as the number of observations tends to infinity. Therefore, this model selection method is useful for large models because it allows insignificant variables to be identified directly without the need for hypothesis testing. The adaptive LASSO expectile method on the all observations produces also estimators that are shrunk directly to zero. However, cross-validation on a large validation set further enhances the robustness of our model estimator.  This result is particularly noteworthy in the context of numerical analysis, as the \textit{ernet} function in the R \textit{SALES} package does not provide hypothesis tests on the parameters of a linear model. Furthermore, the CV expectile  method is indicated when model errors are asymmetric, thereby precluding the classic assumption of normality therefore renders the use of the classical least squares method inappropriate. The simulations presented in the previous section showed that, compared to the CV quantile method, which can also be used in the case of asymmetric errors, the CV expectile  method is more accurate and has a shorter execution time, especially for a model with a large number of explanatory variables. Another very important aspect obtained by the simulations is that  when non-zero  values of $\beta^*_j$ are close to zero, the CV expectile estimator identifies irrelevant variables very well,  unlike the CV LS, CV quantile, and adaptive LASSO methods.
\section{Proofs of main results}
\label{section_proofs}
In this section, we present the proofs of the three theorems stated in Section \ref{section_Model}.\\
The proof of each theorem will be established by demonstrating the following two relations for any over-specified model, ${\cal M}_o$, and any under-specified model,  ${\cal M}_u$:
\begin{equation}
	\label{rel_Mo}
	\lim_{n  \rightarrow \infty} \PP[CVS({\cal M}_o )>CVS({\cal M}^*)]=1
\end{equation}
and
\begin{equation}
	\label{rel_Mu}
	\lim_{n  \rightarrow \infty} \PP[CVS({\cal M}_u) >CVS({\cal M}^*)]=1.
\end{equation}	
For proving (\ref{rel_Mo}), we will decompose $CVS({\cal M}_o) -CVS({\cal M}^*)$ into:
\begin{align}
		CVS({\cal M}_o) -CVS({\cal M}^*) & = \frac{1}{n_{\cal V}}\sum_{i \in {\cal V}} \rho_\tau (Y_i -\eX^\top_{i,{\cal M}_o} \widehat \eb_{n_{\cal A},{\cal M}_o}) - \frac{1}{n_{\cal V}} \sum_{i \in {\cal V} }\rho_\tau (Y_i -\eX^\top_{i,{\cal M}^*} \widehat \eb_{n_{\cal A},{\cal M}^*}) \nonumber \\
		& =\frac{1}{n_{\cal V}}\sum_{i \in {\cal V}} \bigg( \rho_\tau (Y_i -\eX^\top_{i,{\cal M}_o} \widehat \eb_{n_{\cal A},{\cal M}_o}) -\rho_\tau(\varepsilon_i)\bigg) \ \nonumber \\
		& \quad -  \frac{1}{n_{\cal V}} \sum_{i \in {\cal V} } \bigg(\rho_\tau (Y_i -\eX^\top_{i,{\cal M}^*} \widehat \eb_{n_{\cal A},{\cal M}^*})-\rho_\tau(\varepsilon_i)\bigg) \nonumber \\
		& \equiv A_n({\cal M}_o) - B_n({\cal M}^*).
	\label{u1}
\end{align}
For proving relation (\ref{rel_Mu}), we will study the decomposition:
\begin{equation}
CVS({\cal M}_u) -CVS({\cal M}^*) \equiv A_n({\cal M}_u)- B_n({\cal M}^*).
\label{rel_Mub}
\end{equation}
To study $A_n({\cal M}_o)$, we write it under the form:
\begin{align}
		A_n( {\cal M}_o) & =\frac{1}{ n_{\cal V}}  \sum_{i  \in {\cal V}} \big(\rho_\tau(\varepsilon_i - \eX_{i,{\cal M}_o}^\top  \frac{\widehat \ed_{{n_{\cal A},{\cal M}}_o}}{r_{ n_{\cal A}}} )- \rho_\tau(\varepsilon_i)\big) \nonumber\\
		& =\bigg(\frac{1}{ n_{\cal V}}  \sum_{i  \in {\cal V}}  \big(\rho_\tau(\varepsilon_i - \eX_{i,{\cal M}_o}^\top  \frac{\widehat \ed_{{n_{\cal A},{\cal M}}_o}}{r_{ n_{\cal A}}}) - \rho_\tau(\varepsilon_i)\big)  \nonumber\\
		& \qquad - \frac{1}{ n_{\cal V}}  \sum_{i  \in {\cal V}}  \eE \big[\rho_\tau(\varepsilon_i - \eX_{i,{\cal M}_o}^\top  \frac{\widehat \ed_{{n_{\cal A},{\cal M}}_o}}{r_{ n_{\cal A}}}) - \rho_\tau(\varepsilon_i)\big]   \bigg) \nonumber \\
		& +\frac{1}{ n_{\cal V}}  \sum_{i  \in {\cal V}}  \eE \big[\rho_\tau(\varepsilon_i - \eX_{i,{\cal M}_o}^\top  \frac{\widehat \ed_{{n_{\cal A},{\cal M}}_o}}{r_{ n_{\cal A}}}) - \rho_\tau(\varepsilon_i)\big]   \equiv \Delta_1 +\Delta_2.
	\label{u2}
\end{align}
\noindent  {\bf Proof of Theorem \ref{Theorem_pfix}}.  \\
\underline{We begin by showing  relation (\ref{rel_Mo}).}\\
By a Taylor expansion, we have for $t \rightarrow 0$: $\eE[\rho_\tau(\varepsilon -t)-\rho_\tau(\varepsilon)]=-\eE[g_\tau(\varepsilon)]t+2^{-1} \eE[h_\tau(\varepsilon)]t^2+o(t^2)=2^{-1} \eE[h_\tau(\varepsilon)]t^2+o(t^2).$ On the other hand, using assumption (A1) from where we have that $(\varepsilon_i)_{i \in {\cal A}}$ are  independent of $(\varepsilon_i)_{i \in {\cal V}}$ we deduce that $(\varepsilon_i)_{i \in {\cal V}}$  are independent of  $\widehat \ed_{{n_{\cal A},{\cal M}}_o}$.  Then, taking into account    assumption (A2) and the   convergence rate  of $\widehat{\eb}_{n_{\cal A},{\cal M}_o}$, we obtain that: 
\begin{equation}
	\Delta_2 =\frac{1}{ 2 n_{\cal V}}  \sum_{i  \in {\cal V}}  \eE[h_\tau(\varepsilon)] \frac{\eE\big[\big(\eX^\top_{i,{\cal M}_o} \widehat \ed_{{n_{\cal A},{\cal M}}_o}\big)^2\big]}{r^2_{n_{\cal A}}}\big(1+o_\PP(1)\big)=O \bigg( \frac{\big\|\widehat \ed_{{n_{\cal A},{\cal M}}_o}\big\|_2^2}{r^2_{n_{\cal A}}}\bigg)=O(r^{-2}_{n_{\cal A}}).
	\label{Delta2}
\end{equation}
In order to study $\Delta_1$, we write it in the form:
\begin{equation}
	\Delta_1 =\frac{1}{ n_{\cal V}}  \sum_{i  \in {\cal V}}  \bigg(-\frac{\widehat \ed^\top_{{n_{\cal A},{\cal M}}_o}}{r_{n_{\cal A}}}g_\tau(\varepsilon_i)\eX_{i,{\cal M}_o} +D_{1,i}-\eE[D_{1,i}] \bigg),
	\label{Delta1}
\end{equation}
with 
\[
D_{1,i} \equiv  \rho_\tau(\varepsilon_i - \eX_{i,{\cal M}_o}^\top  \frac{\widehat \ed_{{n_{\cal A},{\cal M}}_o}}{r_{ n_{\cal A}}}) - \rho_\tau(\varepsilon_i) +  \frac{\widehat \ed^\top_{{n_{\cal A},{\cal M}}_o}}{r_{ n_{\cal A}}}g_\tau(\varepsilon_i) \eX_{i,{\cal M}_o}.
\]
For the first term on the right-hand side of relation (\ref{Delta1}), using assumptions (A1) and (A2) together with $r_n=n_{\cal A}^{1/2}$, we have by the central limit theorem (CLT):
\begin{equation}
	\frac{1}{ n_{\cal V}}  \sum_{i  \in {\cal V}}   \frac{\widehat \ed^\top_{{n_{\cal A},{\cal M}}_o}}{r_{ n_{\cal A}}}g_\tau(\varepsilon_i) \eX_{i,{\cal M}_o} =O_\PP \bigg(\frac{\|\widehat \ed_{{n_{\cal A},{\cal M}}_o}\|_2}{ n^{1/2}_{\cal V}  n^{1/2}_{\cal A}}\bigg) =O_\PP(n_{\cal A}^{-1/2} n_{\cal V}^{-1/2}).
	\label{dd}
\end{equation}
In order to analyze the last two terms on the right-hand side of (\ref{Delta1}), Lemma 1 from \cite{Gu-Zou.16} and the Cauchy–Schwarz inequality are applied. We then  obtain, with probability 1:
\begin{equation}
	\label{ineg_D1i}
	|D_{1,i}|=\frac{1}{r^2_{n_{\cal A}}}\big| \eX_{i,{\cal M}_o}^\top \widehat \ed_{{n_{\cal A},{\cal M}}_o} \big|^2 V_i \leq \|\eX_{i,{\cal M}_o} \|^2_2   \| \widehat \eb_{n_{\cal A},{\cal M}_o} - \eb^*_{{\cal M}_o}\|^2_2V_i,
\end{equation}
where $V_i$ is a random variable with values in the interval $[\min(\tau, 1-\tau),\max(\tau, 1-\tau)]$.  
Then, using assumptions (A1) and (A2) and the fact that the random variable $V_i$ takes values belonging to the interval $(0,1)$, we obtain:
\begin{equation*}
	\begin{split} 
		\frac{1}{ n_{\cal V}}  \sum_{i  \in {\cal V}}  \big(D_{1,i}-\eE[D_{1,i}]\big) & = \frac{1}{ n_{\cal V}}  O_\PP \bigg( \sqrt{\sum_{i  \in {\cal V}} \Var [D_{1,i}]}\bigg) \leq   \frac{1}{ n_{\cal V}}  O_\PP \bigg( \sqrt{\sum_{i  \in {\cal V}} \eE [D^2_{1,i}]}\bigg) \\
		& \leq  \frac{1}{ n_{\cal V}}  O_\PP \bigg( \sqrt{\sum_{i  \in {\cal V}} \eE [\|\eX_{i,{\cal M}_o} \|^2_2   \| \widehat \eb_{n_{\cal A},{\cal M}_o} - \eb^*_{{\cal M}_o}\|^2_2V_i ]^2}\bigg)\\
		&=  \frac{1}{ n_{\cal V}}  O_\PP \bigg( \sqrt{n_{\cal V} \eE \big[ \| \widehat \eb_{n_{\cal A},{\cal M}_o} - \eb^*_{{\cal M}_o}\|^4_2} \big]\bigg)=O_\PP\big(n_{\cal V}^{-1/2}\eE\big[\| \widehat \eb_{n_{\cal A},{\cal M}_o} - \eb^*_{{\cal M}_o}\|^2_2\big]\big)\\
		& =O_\PP(n_{\cal V}^{-1/2} n^{-1}_{\cal A}).
	\end{split}
\end{equation*}
Combining this last relation  with relations (\ref{Delta2}) and (\ref{dd}), we obtain that: $\Delta_1 = O_\PP(n_{\cal V}^{-1/2} n^{-1/2}_{\cal A})$.  The order of $\Delta_1$ combined with  relation (\ref{Delta2}), implies that: $\Delta_2/ \Delta_1=O_\PP (n_{\cal V}^{1/2} n^{-1/2}_{\cal A})$. Thus, since $n_{\cal V} \gg n_{\cal A}$ by assumption (A4), we deduce that $\Delta_2 \gg |\Delta_1|$ with probability converging to 1 and then 
\begin{equation}
	A_n({\cal M}_o)=\Delta_2 (1+o_\PP(1))=O_\PP(n^{-1}_{\cal A}).
	\label{u3}
\end{equation}
 We now study $B_n({\cal M}^*)$. For any $i \in {\cal V}$, we have:
\[
Y_i - \eX_{i,{\cal M}^*}^\top \widehat \eb_{n_{\cal A},{\cal M}^*}=\varepsilon_i - \eX_{i,{\cal M}^*}^\top  \big( \widehat \eb_{n_{\cal A},{\cal M}^*} - \eb^*_{{\cal M}^*} \big)= \varepsilon_i - \eX_{i,{\cal M}^*}^\top \frac{ \widehat \ed^\top_{n_{\cal A},{\cal M}^*}}{r_{n_{\cal A}}}.
\]
Similarly, using also the fact that $\eE[h_\tau(\varepsilon)] >0$, we show that:
\begin{equation}
	\label{Bn}
	B_n({\cal M}^*)=\frac{1}{n_{\cal V}}  \sum_{i  \in {\cal V}}\eE[h_\tau(\varepsilon_i)]	\frac{\big(\eX^\top_{i,{\cal M}^*} \widehat \ed_{{n_{\cal A},{\cal M}^*}}\big)^2}{r^2_{n_{\cal A}}}\big(1+o_\PP(1)\big)=O_\PP\bigg(\frac{\big\|\ed_{{n_{\cal A},{\cal M}^*}}\big\|_2^2}{r^2_{n_{\cal A}}}\bigg)=O_\PP(n^{-1}_{\cal A})>0.
\end{equation}
Then, by relations (\ref{u1}), (\ref{u2}), (\ref{u3}) and (\ref{Bn}), we have:
\begin{equation*}
	\begin{split} 
		A_n({\cal M}_o)	-B_n({\cal M}^*)=&\bigg( \frac{1}{n_{\cal V}}  \sum_{i  \in {\cal V}}\eE[h_\tau(\varepsilon_i)]\widehat \ed_{{n_{\cal A},{\cal M}}_o} \frac{ \eX^\top_{i,{\cal M}_o} \eX_{i,{\cal M}_o} }{r^2_{n_{\cal A}}} \widehat \ed^\top_{{n_{\cal A},{\cal M}}_o}\\
		&-\frac{1}{n_{\cal V}}  \sum_{i  \in {\cal V}}\eE[h_\tau(\varepsilon_i)]\widehat \ed_{{n_{\cal A},{\cal M}^*}} \frac{ \eX^\top_{i,{\cal M}^*} \eX_{i,{\cal M}^*} }{r^2_{n_{\cal A}}} \widehat \ed^\top_{{n_{\cal A},{\cal M}^*}}\bigg) (1+o_\PP(1)).
	\end{split} 
\end{equation*}
For a model ${\cal M}$, let the square matrix of dimension $|{\cal M}|$: 
$$\eSS_{\cal M} \equiv \lim_{n  \rightarrow \infty} n^{-1}_{\cal V} \sum_{i  \in {\cal V}}\eE[h_\tau(\varepsilon_i)]  \eX^\top_{i,{\cal M}}  \eX_{i,{\cal M}}.$$
Then, by Slutsky’s lemma together with  the independence of model errors  $(\varepsilon_i)_{i \in {\cal V}}$ from $\widehat \ed^\top_{{n_{\cal A},{\cal M}}_o}$ and from $\widehat \ed^\top_{{n_{\cal A},{\cal M}}^*}$, we obtain:
\[
A_n({\cal M}_o)	-B_n({\cal M}^*) \overset{\cal L} {\underset{n \rightarrow \infty}{\longrightarrow}}\ed^\top_{{\cal M}_o}\eSS_{{\cal M}_o}\ed_{{\cal M}_o} -\ed^\top_{{\cal M}^*}\eSS_{{\cal M}^*}\ed_{{\cal M}^*} >0.
\]
The limit distribution is strictly positive because ${\cal M}^* \subsetneq {\cal M}_o$. The proof of relation (\ref{rel_Mo}) is complete.\\
\underline{We now establish  relation (\ref{rel_Mu}).}\\
By relation (\ref{Bn}), we have $ B_n({\cal M}^*)=O_\PP(n^{-1}_{\cal A})>0$. Concerning 
\[
A_n( {\cal M}_u)  =\frac{1}{ n_{\cal V}}  \sum_{i  \in {\cal V}} \bigg(\rho_\tau(Y_i - \eX_{i,{\cal M}_u}^\top  \widehat \eb_{n_{\cal A},{\cal M}_u})- \rho_\tau(\varepsilon_i)\bigg)
\]
we write $\widehat \eb_{n_{\cal A},{\cal M}_u}$ under the form $r_{n_{\cal A}}^{-1}\widehat \ed_{{n_{\cal A},{\cal M}}_u} +\eb^*_{{\cal M}_u} $, which implies:
\begin{equation*}
	A_n( {\cal M}_u)  =\frac{1}{ n_{\cal V}}  \sum_{i  \in {\cal V}} \bigg(\rho_\tau(Y_i - r_{n_{\cal A}}^{-1} \eX_{i,{\cal M}_u}^\top  \widehat \ed_{{n_{\cal A},{\cal M}}_u}- \eX_{i,{\cal M}_u}^\top\eb^*_{{\cal M}_u} )- \rho_\tau(\varepsilon_i)\bigg).
\end{equation*}
In this case, for the model ${\cal M}_u $ we obtain the consistent estimators $\widehat \eb_{n_{\cal A},{\cal M}_u}$ of $\eb^*_{{\cal M}_u}$ with a convergence rate of order $r_{n_{\cal A}}^{-1}$. Then, $ r_{n_{\cal A}}^{-1}\widehat \ed_{{n_{\cal A},{\cal M}}_u}  =O_\PP(r_{n_{\cal A}}^{-1})$.\\
We write $ \rho_\tau(Y_i -  r_{n_{\cal A}}^{-1}\eX_{i,{\cal M}_u}^\top {\widehat \ed_{{n_{\cal A},{\cal M}}_u}} - \eX_{i,{\cal M}_u}^\top\eb^*_{{\cal M}_u} )$ as:
\begin{equation}
	\label{uu}
	\rho_\tau(Y_i -  r_{n_{\cal A}}^{-1}\eX_{i,{\cal M}_u}^\top \widehat \ed_{{n_{\cal A},{\cal M}}_u}- \eX_{i,{\cal M}_u}^\top\eb^*_{{\cal M}_u} )-\rho_\tau(Y_i - \eX_{i,{\cal M}_u}^\top\eb^*_{{\cal M}_u} )+\rho_\tau(Y_i - \eX_{i,{\cal M}_u}^\top\eb^*_{{\cal M}_u} ).
\end{equation}
Using Lemma 1 of \cite{Gu-Zou.16}, we have, for any $i \in {\cal V}$:
\begin{align}
		&\rho_\tau(Y_i -  r_{n_{\cal A}}^{-1}\eX_{i,{\cal M}_u}^\top   \widehat \ed_{{n_{\cal A},{\cal M}}_u} - \eX_{i,{\cal M}_u}^\top\eb^*_{{\cal M}_u} )-\rho_\tau(Y_i - \eX_{i,{\cal M}_u}^\top\eb^*_{{\cal M}_u} ) \nonumber  \\
		&= r_{n_{\cal A}}^{-1}g_\tau(Y_i - \eX_{i,{\cal M}_u}^\top\eb^*_{{\cal M}_u} )  \eX_{i,{\cal M}_u}^\top \widehat \ed_{{n_{\cal A},{\cal M}}_u}+O_\PP\big( r_{n_{\cal A}}^{-1}\eX_{i,{\cal M}_u}^\top  \widehat \ed_{{n_{\cal A},{\cal M}}_u}\big)^2,
	\label{lp} 
\end{align}
for which,   applying the same reasoning as in the case of (\ref{ineg_D1i}), it holds that $O_\PP\big( r_{n_{\cal A}}^{-1}\eX_{i,{\cal M}_u}^\top   \widehat \ed_{{n_{\cal A},{\cal M}}_u} \big)^2$ $=O_\PP( r_{n_{\cal A}}^{-2})=O_\PP(n^{-1}_{\cal A})$.\\
We study $ n_{\cal V}^{-1}  \sum_{i  \in {\cal V}} r_{n_{\cal A}}^{-1}g_\tau(Y_i - \eX_{i,{\cal M}_u}^\top\eb^*_{{\cal M}_u} )  \eX_{i,{\cal M}_u}^\top \widehat \ed_{{n_{\cal A},{\cal M}}_u} $. Because ${\cal M}^* \nsubseteq {\cal M}_u$ then $Y_i=\eX_{i,{\cal M}_u}^\top\eb^*_{{\cal M}_u} +\eX_{i,{\cal M}^* \setminus {\cal M}_u}^\top\eb^*_{{\cal M}^* \setminus {\cal M}_u} +\varepsilon_i$, which implies that:
\begin{equation*}
	 n^{-1}_{\cal V}  \sum_{i  \in {\cal V}} r_{n_{\cal A}}^{-1}g_\tau(Y_i - \eX_{i,{\cal M}_u}^\top\eb^*_{{\cal M}_u} )  \eX_{i,{\cal M}_u}^\top \widehat \ed_{{n_{\cal A},{\cal M}}_u} 
\end{equation*}
\begin{equation} 
	\begin{split}
		&	 =n^{-1}_{\cal V} \sum_{i  \in {\cal V}}  r_{n_{\cal A}}^{-1}\bigg(g_\tau(\varepsilon_i + \eX_{i,{\cal M}^* \setminus {\cal M}_u}^\top\eb^*_{{\cal M}^* \setminus {\cal M}_u} )-g_\tau(\varepsilon_i) \bigg) \eX_{i,{\cal M}_u}^\top \widehat \ed_{{n_{\cal A},{\cal M}}_u}  \\
		&  \qquad + n^{-1}_{\cal V}  \sum_{i  \in {\cal V}}  r_{n_{\cal A}}^{-1} g_\tau(\varepsilon_i) \eX_{i,{\cal M}_u}^\top \widehat \ed_{{n_{\cal A},{\cal M}}_u} 
	\end{split}
	\label{eq5}
\end{equation} 
The last term of  relation (\ref{eq5}) is $O_\PP(n_{\cal A}^{-1/2} n_{\cal V}^{-1/2})$ by CLT.\\
We now use Lemma 2 from \cite{Gu-Zou.16} and can write:
\begin{align*}
	&n^{-1}_{\cal V}  \sum_{i  \in {\cal V}}r_{n_{\cal A}}^{-1} \bigg(g_\tau(\varepsilon_i + \eX_{i,{\cal M}^* \setminus {\cal M}_u}^\top\eb^*_{{\cal M}^* \setminus {\cal M}_u} )-g_\tau(\varepsilon_i) \bigg) \eX_{i,{\cal M}_u}^\top \widehat \ed_{{n_{\cal A},{\cal M}}_u} \\
	& \qquad =	n^{-1}_{\cal V}  \sum_{i  \in {\cal V}} r_{n_{\cal A}}^{-1} \eX_{i,{\cal M}^* \setminus {\cal M}_u}^\top\eb^*_{{\cal M}^* \setminus {\cal M}_u}  \eX_{i,{\cal M}_u}^\top \widehat \ed_{{n_{\cal A},{\cal M}}_u} \widetilde V_i \\
	&  \qquad= O_\PP(\| r^{-1}_{n_{\cal A}}\widehat \ed_{n_{\cal A},{\cal M}_u}  \|_2)=O_\PP(r_{n_{\cal A}}^{-1})= O_\PP(n^{-1/2}_{\cal A}),
\end{align*}
where $\widetilde V_i $ is a random variable taking values in the interval $[-2\max(\tau, 1-\tau), 2\max(\tau, 1-\tau)]$.\\
Taking into account relations (\ref{uu}), (\ref{lp}), (\ref{eq5}), we obtain that $ A_n( {\cal M}_u) $ becomes
\[
A_n( {\cal M}_u)  =\frac{1}{ n_{\cal V}}  \sum_{i  \in {\cal V}} \big(\rho_\tau(Y_i -\eX_{i,{\cal M}_u}^\top\eb^*_{{\cal M}_u} ) - \rho_\tau(\varepsilon_i)\big) +O_\PP(n^{-1/2}_{\cal A}).
\]
However, for any $i \in {\cal V}$, we can write $\varepsilon_i=Y_i - \eX_i^\top \eb^*_{{\cal M}^*}=Y_i -\eX_{i,{\cal M}_u}^\top\eb^*_{{\cal M}_u} -\eX_{i,{\cal M}^* \setminus  {\cal M}_u}^\top\eb^*_{{\cal M}^* \setminus{\cal M}_u}$ and we rely on assumption (A3).\\ 
Then, for large enough $n$, we obtain that $CVS({\cal M}_u)- CVS({\cal M}^*)>0$ , from which relation (\ref{rel_Mu}) follows. 
\hspace*{\fill}$\blacksquare$ \\

\noindent  {\bf Proof of Theorem \ref{Theorem_pn1}}.  \\
We begin by establishing relation (\ref{rel_Mo}), using the decomposition $	CVS({\cal M}_o) -CVS({\cal M}^*) \equiv A_n({\cal M}_o) - B_n({\cal M}^*)$. \\
We first study $A_n({\cal M}_o)  \equiv  \Delta_1+\Delta_2$. For $\Delta_2$, following a reasoning similar to that in the proof of Theorem \ref{Theorem_pfix} and using assumption (A2), we obtain: 
\begin{equation*}
		\Delta_2= \frac{1}{n_{\cal V}} \sum_{i\in {\cal V}} \eE \big[\rho_\tau(\varepsilon_i- \eX^\top_{{i,{\cal M}}_o} \frac{\widehat \ed_{n_{\cal A},{\cal M}_o}}{r_{n_{\cal A}} }) - \rho_\tau(\varepsilon_i)\big]
		 =O_\PP \bigg(\frac{\|\widehat \ed_{n_{\cal A},{\cal M}_o} \|^2_2}{r^2_{n_{\cal A}} }\bigg)=O_\PP(r^{-2}_{n_{\cal A}})=O_\PP \bigg(\frac{p_n}{n_{\cal A}}\bigg).
\end{equation*}
For the last relation, we used Theorem 2.1 of \cite{Ciuperca.21} together with assumptions (A1), (A2), (A5).\\
The term  $\Delta_1 \equiv n^{-1}_{\cal V} \sum_{i\in {\cal V}} \big( \rho_\tau(\varepsilon_i -  \eX^\top_{{i,{\cal M}}_o} r^{-1}_{n_{\cal A}}   \widehat \ed_{n_{\cal A},{\cal M}_o}  ) -\rho_\tau(\varepsilon_i) \big)-n^{-1}_{\cal V} \sum_{i\in {\cal V}} \eE\big[ \rho_\tau(\varepsilon_i -  \eX^\top_{{i,{\cal M}}_o} r^{-1}_{n_{\cal A}}   \widehat \ed_{n_{\cal A},{\cal M}_o}  ) -\rho_\tau(\varepsilon_i) \big]$ can be written as in (\ref{Delta1}). \\
Using assumption (A1), we obtain: $\eE[g_\tau(\varepsilon_i)  \eX^\top_{{i,{\cal M}}_o} \widehat \ed_{n_{\cal A},{\cal M}_o} ]=\oo_{p_n}$ and using assumption  (A2) we have: 
$
\Var[\sum_{i\in {\cal V}} g_\tau(\varepsilon_i)  \eX^\top_{{i,{\cal M}}_o} \widehat \ed_{n_{\cal A},{\cal M}_o}]=O_\PP(n_{\cal V} \|  \widehat \ed_{n_{\cal A},{\cal M}_o}\|_2^2)$. 
Then, using CLT, we obtain:
\begin{equation}
	\frac{1}{n_{\cal V}} \sum_{i\in {\cal V}} \frac{ \widehat \ed^\top_{n_{\cal A},{\cal M}_o} }{r_{n_{\cal A}}} g_\tau(\varepsilon_i) \eX_{{i,{\cal M}}_o}=O_\PP(n_{\cal V}^{-1/2}\| \widehat \eb_{n_{\cal A},{\cal M}_o}- \eb^*_{{\cal M}_o}\|_2)=O_\PP(n_{\cal V}^{-1/2}r_{n_{\cal A}}^{-1}).
	\label{Au}
\end{equation}
On the one hand, we have: $ 
 n^{-1}_{\cal V} \sum_{i\in {\cal V}} \big(D_{1,i}-\eE[D_{1,i}]\big)\leq  n^{-1}_{\cal V} O_\PP \bigg(\sqrt{\sum_{i\in {\cal V}} \eE[D^2_{1,i}]}\bigg)$.
Furthermore, using Lemma 1 of \cite{Gu-Zou.16}, we obtain:
\begin{equation*}
	\begin{split} 
		D_{1,i}& =\frac{1}{r^2_{n_{\cal A}}} \big| \widehat \ed^\top_{n_{\cal A},{\cal M}_o} \eX_{{i,{\cal M}}_o} \big|^2 V_i , \quad \textrm{with random variable } V_i   \in [\min(\tau, 1-\tau),  \max(\tau, 1-\tau)]\\
		& \leq \|\eX_{{i,{\cal M}}_o}  \|^2_2   \| \widehat \eb_{n_{\cal A},{\cal M}_o}- \eb^*_{{\cal M}_o}\|_2^2 , \qquad \textrm{with probability 1,}\\
		&\leq  r^{-2}_{n_{\cal A}} \max_{i \in  {\cal V}} \| \eX_{i,{\cal M}_o }\|^2_2.
	\end{split}
\end{equation*}
The issue arises from the fourth power in $ \big| \widehat \ed^\top_{n_{\cal A},{\cal M}_o} \eX_{{i,{\cal M}}_o} \big|^4$ , which prevents applying assumption (A2) when considering the sum $ \widehat \ed^\top_{n_{\cal A},{\cal M}_o} \sum_{i \in {\cal V}} \eX_{{i,{\cal M}}_o}  \eX^\top_{{i,{\cal M}}_o} \widehat \ed_{n_{\cal A},{\cal M}_o} $.  Therefore, 
\[
\frac{1}{n_{\cal V}}\sqrt{\sum_{i \in {\cal V}}\eE[D^2_{1,i}]} \leq \frac{1}{n_{\cal V}} \sqrt{r^{-4}_{{n_{\cal A}}}\sum_{i \in {\cal V}} \max_{j \in  {\cal V}} \| \eX_j\|^4_2 }=r^{-2}_{n_{{\cal A}}} \frac{ \max_{j \in  {\cal V}} \| \eX_j\|^2_2 }{n^{1/2}_{\cal V}},
\]
from where
\begin{equation}
	\label{Bu}
	\frac{1}{n_{\cal V}} \sum_{i\in {\cal V}} \big(D_{1,i}-\eE[D_{1,i}]\big) \leq O_\PP \bigg(r^{-2}_{n_{{\cal A}}} \frac{ \max_{j \in  {\cal V}} \| \eX_j\|^2_2 }{n^{1/2}_{\cal V}}\bigg).
\end{equation}
Thus, taking into account relations (\ref{Au}) and (\ref{Bu}), we have for $\Delta_1$ that:
\[
\Delta_1 = O_\PP \bigg( \max  \bigg(r^{-2}_{n_{{\cal A}}} \frac{ \max_{j \in  {\cal V}} \| \eX_j\|^2_2 }{n^{1/2}_{\cal V}}, n^{-1/2}_{\cal V}r^{-1}_{n_{\cal A}}\bigg)  \bigg).
\]
We compare $\Delta_1$ and $\Delta_2$. Using assumption (A6), we obtain:
\begin{equation*}
r^{-2}_{{n_{\cal A}}}	\frac{ \max_{j \in  {\cal V}} \| \eX_j\|^2_2 }{ n^{1/2}_{\cal V}} 	r^{2}_{n_{{\cal A}}} 	=\frac{ \max_{j \in  {\cal V}} \| \eX_j\|^2_2 }{ n^{1/2}_{\cal V}} {\underset{n \rightarrow \infty}{\longrightarrow}} 0 .
\end{equation*}
By the fact that $n_{\cal A} \ll n_{\cal V}$ of assumption (A4), we obtain that
\begin{equation*}
	n^{-1/2}_{\cal V}r^{-1}_{n_{\cal A}} r^{2}_{n_{{\cal A}}} 	=n^{-1/2}_{\cal V}r_{n_{\cal A}} =\bigg(\frac{n_{\cal A}}{p_n}\bigg)^{1/2} \frac{1}{n^{1/2}_{\cal V}} \ll \frac{n^{1/2}_{\cal V}}{p^{1/2}_n} \frac{1}{n^{1/2}_{\cal V}}{\underset{n \rightarrow \infty}{\longrightarrow}} 0 .
\end{equation*}
Then we obtain that  $\Delta_2 \gg \Delta_1$, which implies 
\[
A_n({\cal M}_o)=\Delta_2\big(1+o_\PP(1)\big)=O_\PP \big(r^{-2}_{n_{\cal A}}\big).
\]
We now study $B_n({\cal M}^*)$. Similarly, as for relation (\ref{Bn}) in the proof of Theorem \ref{Theorem_pfix}, using assumption (A2), we can write:
\begin{align}
	B_n({\cal M}^*) & =O_\PP\bigg(\frac{\| \widehat \ed_{n_{\cal A},{\cal M}^*}  \|_2^2}{r^2_{n_{\cal A}}}\bigg)=O_\PP\big(\|\widehat \eb_{n_{\cal A},{\cal M}^*} - \eb^*_{{\cal M}^*}  \|_2^2\big)=O_\PP\big(r^{-2}_{n_{\cal A}}\big)	.
	\label{Dnpn}
\end{align}
In conclusion:
\begin{equation*}
	\begin{split} 
		r^2_{n_{\cal A}} \big(A_n({\cal M}_o) &-B_n({\cal M}^*) \big)- \frac{1}{n_{\cal V}} \sum_{i\in {\cal V} } \eE[h_\tau(\varepsilon_i)] \bigg( \widehat \ed_{n_{\cal A},{\cal M}_o} \eX^\top_{{i,{\cal M}}_o} \eX_{{i,{\cal M}}_o}  \widehat \ed^\top_{n_{\cal A},{\cal M}_o}  \\
		& -  \widehat \ed_{n_{\cal A},{\cal M}^*} \eX^\top_{{i,{\cal M}}^*} \eX_{{i,{\cal M}}^*}  \widehat \ed^\top_{n_{\cal A},{\cal M}^*}\bigg) \overset{\PP} {\underset{n \rightarrow \infty}{\longrightarrow}} >0, \qquad \textrm{since } {\cal M}^* \nsubseteq {\cal M}_o
	\end{split}
\end{equation*}
and relation (\ref{rel_Mo}) is proved.\\
We now prove relation (\ref{rel_Mu}), for which we study  decomposition (\ref{rel_Mub}) taking into account (\ref{u2}).  By  relation (\ref{Dnpn}) we have $B_n({\cal M}^*) =O_\PP\big(r^{-2}_{ n_{\cal A}}\big) $.\\
To study $A_n({\cal M}_u)$, we consider   the decomposition $\widehat \eb_{n_{\cal A},{\cal M}_u}=\big(\widehat \eb_{n_{\cal A},{\cal M}_u} - \eb^*_{{\cal M}_u}\big)+\eb^*_{{\cal M}_u}$. Therefore, we obtain:
\[
CVS({\cal M}_u) =	\frac{1}{n_{\cal V}} \sum_{i\in {\cal V} } \bigg(\rho_\tau(Y_i - \eX_{i,{\cal M}_u}^\top  \frac{\widehat \ed_{{n_{\cal A},{\cal M}}_u}}{r_{n_{\cal A}}}- \eX_{i,{\cal M}_u}^\top\eb^*_{{\cal M}_u} )- \rho_\tau(\varepsilon_i)\bigg) .
\]
We write:
\begin{equation*}
	\begin{split} 
		&\rho_\tau(Y_i - \eX_{i,{\cal M}_u}^\top  \frac{\widehat \ed_{{n_{\cal A},{\cal M}}_u}}{r_{n_{\cal A}}}- \eX_{i,{\cal M}_u}^\top\eb^*_{{\cal M}_u} )\\
		&=\rho_\tau(Y_i - \eX_{i,{\cal M}_u}^\top  \frac{\widehat \ed_{{n_{\cal A},{\cal M}}_u}}{r_{n_{\cal A}}}- \eX_{i,{\cal M}_u}^\top\eb^*_{{\cal M}_u} )-\rho_\tau(Y_i  - \eX_{i,{\cal M}_u}^\top\eb^*_{{\cal M}_u} )	 +\rho_\tau(Y_i  - \eX_{i,{\cal M}_u}^\top\eb^*_{{\cal M}_u} )	
	\end{split}
\end{equation*}
Thus, following the proof of Theorem \ref{Theorem_pfix}, we study:
\begin{align*}
	& 	\frac{1}{n_{\cal V}} \sum_{i\in {\cal V} } \bigg(\rho_\tau(Y_i - \eX_{i,{\cal M}_u}^\top  \frac{\widehat \ed_{{n_{\cal A},{\cal M}}_u}}{r_{n_{\cal A}}}- \eX_{i,{\cal M}_u}^\top\eb^*_{{\cal M}_u} )-\rho_\tau(Y_i  - \eX_{i,{\cal M}_u}^\top\eb^*_{{\cal M}_u} )\bigg).
\end{align*}
Under assumptions (A1), (A2), (A5), the   convergence rate of $\widehat \eb_{n_{\cal A},{\cal M}_u}$ toward $\eb^*_{{\cal M}_u}$ is $r^{-1}_{n_{\cal A}}$, that is: $\| \widehat \eb_{n_{\cal A},{\cal M}_u} -\eb^*_{{\cal M}_u} \|_2=O_\PP(r^{-1}_{n_{\cal A}})$.  Then, for relation (\ref{lp}), we have:
\[
O_\PP \big( r^{-1}_{n_{\cal A}} \big| \eX^\top_{i,{\cal M}_u} \widehat \ed_{{n_{\cal A},{\cal M}}_u}\big|\big) \leq O_\PP \big(  r^{-1}_{n_{\cal A}} \| \eX_{i,{\cal M}_u}\|_2 \| \widehat \ed_{{n_{\cal A},{\cal M}}_u}\|_2 \big) \leq O_\PP \big( r^{-1}_{n_{\cal A}} \max_{1 \leqslant i \leqslant n} \| \eX_{i,{\cal M}_u}\|_2 \big)=o_\PP(1).
\]
For the last relation we have used assumption (A5).\\
We now study the following relation for which we have by the Cauchy-Schwarz inequality and Lemma 2 of \cite{Gu-Zou.16} that:
\begin{align}
	&\frac{1}{ n_{\cal V}} \bigg|  \sum_{i  \in {\cal V}}  r_{n_{\cal A}}^{-1}\bigg(g_\tau(\varepsilon_i + \eX_{i,{\cal M}^* \setminus {\cal M}_u}^\top\eb^*_{{\cal M}^* \setminus {\cal M}_u} )-g_\tau(\varepsilon_i) \bigg) \eX_{i,{\cal M}_u}^\top \widehat \ed_{{n_{\cal A},{\cal M}}_u} \bigg| \nonumber \\
	&\leq  \frac{1}{ n_{\cal V}}  \sum_{i  \in {\cal V}}  r_{n_{\cal A}}^{-1} \bigg| g_\tau(\varepsilon_i + \eX_{i,{\cal M}^* \setminus {\cal M}_u}^\top\eb^*_{{\cal M}^* \setminus {\cal M}_u} )-g_\tau(\varepsilon_i)  \bigg| \cdot \bigg|\eX_{i,{\cal M}_u}^\top \widehat \ed_{{n_{\cal A},{\cal M}}_u } \bigg| \nonumber\\
	&\leq  \frac{2}{ n_{\cal V}}  \sum_{i  \in {\cal V}}  r_{n_{\cal A}}^{-1} \bigg|\eX_{i,{\cal M}^* \setminus {\cal M}_u}^\top\eb^*_{{\cal M}^* \setminus {\cal M}_u}\bigg| \cdot \bigg| \eX_{i,{\cal M}_u}^\top \widehat \ed_{{n_{\cal A},{\cal M}}_u}   \bigg| \nonumber \\
	&\leq  \frac{2}{ n_{\cal V}}  \sum_{i  \in {\cal V}}  r_{n_{\cal A}}^{-1} \big\|\eX_{i,{\cal M}^* \setminus {\cal M}_u}\big\|_2  \big\|\eb^*_{{\cal M}^* \setminus {\cal M}_u}\big\|_2   \big\| \eX_{i,{\cal M}_u}\big\|_2  \big\|_2 \widehat \ed_{{n_{\cal A},{\cal M}}_u}   \big\|_2 \nonumber \\
	&\leq   2  r_{n_{\cal A}}^{-1} \max_{i \in {\cal V}} \big\|\eX_{i,{\cal M}^*}\big\|^2_2   \big\|\eb^*_{{\cal M}^* \setminus {\cal M}_u}\big\|_2   \big\|_2 \widehat \ed_{{n_{\cal A},{\cal M}}_u}   \big\|_2 \nonumber \\
	& =o_\PP(1).
	\label{Lu}
\end{align}
For the last relation we have used assumption (A7).
Then, $A_n({\cal M}_u)$ becomes:
\[
A_n({\cal M}_u)=\frac{1}{n_{\cal V}} \sum_{i\in {\cal V} } \bigg(\rho_\tau(Y_i  - \eX_{i,{\cal M}_u}^\top\eb^*_{{\cal M}_u} ) -\rho_\tau(\varepsilon_i)\bigg) +o_\PP(1).
\]
On the other hand, we have the identity $\rho_\tau(\varepsilon_i)=\rho_\tau(Y_i  - \eX_{i,{\cal M}^*}^\top\eb^*_{{\cal M}^*} )=\rho_\tau(Y_i  - \eX_{i,{\cal M}_u}^\top\eb^*_{{\cal M}_u}  - \eX_{i,{\cal M}^* \setminus {\cal M}_u}^\top\eb^*_{{\cal M}^*} \setminus {\cal M}_u)$. Thus, using assumption (A3), we obtain:
\[
A_n({\cal M}_u)=\liminf_{n \rightarrow\infty} k_n+o_\PP(1) >0.
\]
Relation (\ref{rel_Mu}) follows. 
\hspace*{\fill}$\blacksquare$ \\

\noindent  {\bf Proof of Theorem \ref{Theorem_pn2}}.  \\
We only give the elements that are important with respect to the proof of Theorem \ref{Theorem_pn1}.\\
\underline{Proof of  relation (\ref{rel_Mo}).} We have by the proof of Lemma 3.1 of \cite{Ciuperca.21}, that:\\
 $\Var\big[\sum_{i\in {\cal V} } g_\tau(\varepsilon_i)\eX^\top_{{i,{\cal M}}_o} \widehat \ed_{n_{\cal A},{\cal M}_o}\big] =O_\PP\big(n_{\cal V} \|\widehat \ed_{n_{\cal A},{\cal M}_o} \|^2_1\big)$. Using assumptions (A1), (A2), and (A8), then relation (\ref{Au}) becomes, 
\begin{equation}
	\label{Aun}
	\frac{1}{n_{\cal V}} \sum_{i\in {\cal V} } \frac{\eX^\top_{{i,{\cal M}}_o} \widehat \ed_{n_{\cal A},{\cal M}_o}}{r_{n_{\cal A}} }g_\tau(\varepsilon_i)  =O_\PP(n_{\cal V}^{-1/2} \|\widehat \eb_{n_{\cal A},{\cal M}_o} - \eb^*_{{\cal M}_o} \|_1)= O_\PP\big(n_{\cal V}^{-1/2} a_{n_{\cal A}}\big).
\end{equation}
Furthermore, by assumptions (A1), (A2), and (A8), it follows similarly to relation (\ref{ineg_D1i}) that:
\begin{equation*}
		\begin{split}
 D^2_{1,i} & \leq 4 \| \eX^\top_{{i,{\cal M}}_o}\|^4_\infty  \|\widehat \eb_{n_{\cal A},{\cal M}_o} - \eb^*_{{\cal M}_o} \|_1^4 \leq 4 \max_{i \in {\cal V}} \| \eX^\top_{{i,{\cal M}}_o}\|^4_\infty   \|\widehat \eb_{n_{\cal A},{\cal M}_o} - \eb^*_{{\cal M}_o} \|_1^4\\
& = \max_{i \in {\cal V}} \| \eX^\top_{{i,{\cal M}}_o}\|^4_\infty O_\PP( a^4_n). 
 \end{split}
\end{equation*}
  Then, 
\[
n^{-1}_{\cal V}\sqrt{\sum_{i \in {\cal V} } \eE[D^2_{1,i}]} \leq r^{-2}_{n_{\cal A}} \frac{ \max_{j \in {\cal V}} \| \eX^\top_{{j,{\cal M}}_o}\|^2_\infty}{n^{1/2}_{\cal V}}.
\]
Thus,  relation (\ref{Bu}) becomes
\begin{equation}
	\label{Bun}
	\frac{1}{n_{\cal V}} \sum_{i\in {\cal V}} \big(D_{1,i}-\eE[D_{1,i}]\big) \leq \frac{1}{n_{\cal V}}  O_\PP \bigg(r^{-2}_{n_{\cal A}}\frac{ \max_{j \in {\cal V}} \| \eX^\top_{{j,{\cal M}}_o}\|^2_\infty}{ n^{1/2}_{\cal V}}\bigg).
\end{equation}\\
Taking into account relations (\ref{Aun}) and (\ref{Bun}), we have:
\[
\Delta_1=\bigg(-\frac{1}{n_{\cal V}} \sum_{i\in {\cal V} } \frac{\eX^\top_{{i,{\cal M}}_o} \widehat \ed_{n_{\cal A},{\cal M}_o}}{r_{n_{\cal A}} }g_\tau(\varepsilon_i)  \bigg)\big(1+o_\PP(1)\big)=O_\PP \bigg( \max  \bigg(r^{-2}_{n_{{\cal A}}} \frac{ \max_{j \in  {\cal V}} \| \eX_j\|^2_\infty }{n^{1/2}_{\cal V}}, n^{-1/2}_{\cal V}r^{-1}_{n_{\cal A}}\bigg)  \bigg)
\]
and $\Delta_2=O_\PP \big(r^{-2}_{n_{\cal A}} \|\widehat \ed_{n_{\cal A},{\cal M}_o} \|^2_1\big)=O_\PP(r^{-2}_{n_{\cal A}})=O_\PP \big(a^2_{n_{\cal A}}\big)$. \\
Using assumption (A6bis), we obtain, similarly to the case  $0 < c < 1/2$, that  $\Delta_2 \gg \Delta_1$, with probability converging to 1. We therefore deduce $A_n({\cal M}_o)=\Delta_2 \big(1+o_\PP(1)\big)=O_\PP(a^2_{n_{\cal A}})$. We now study  $B_n({\cal M}^*)=O_\PP\big(r^{-2}_{n_{\cal A}} \|\widehat \ed_{n_{\cal A},{\cal M}^*}\|^2_1\big)=O_\PP \big(\| \widehat \eb_{n_{\cal A},{\cal M}^*} -\eb^*_{{\cal M}^*} \|_1^2\big)=O_\PP(a_{n_{\cal A}})$. Similarly, at the end, we have
\[
\PP \big[A_n({\cal M}_o) - B_n({\cal M}^*) >0\big] {\underset{n \rightarrow \infty}{\longrightarrow}} 1.
\]
\underline{Proof of  relation (\ref{rel_Mu}).}\\
We only give the essential relations relevant to the proof of Theorem \ref{Theorem_pn1}. Thus, we have:
\begin{align}
	& 	\frac{1}{n_{\cal V}} \sum_{i\in {\cal V} } \bigg(\rho_\tau(Y_i - \eX_{i,{\cal M}_u}^\top  \frac{\widehat \ed_{{n_{\cal A},{\cal M}}_u}}{r_{n_{\cal A}}}- \eX_{i,{\cal M}_u}^\top\eb^*_{{\cal M}_u} )-\rho_\tau(Y_i  - \eX_{i,{\cal M}_u}^\top\eb^*_{{\cal M}_u} )\bigg) \nonumber \\
	& =	\frac{1}{n_{\cal V}} \sum_{i\in {\cal V} } \eX^\top_{i,{\cal M}^* \setminus {\cal M}_u} \eb^*_{{\cal M}^* \setminus {\cal M}_u}  \eX^\top_{i,{\cal M}_u} \frac{\widehat \ed_{{n_{\cal A},{\cal M}}_u}}{r_{n_{\cal A}}} \widetilde V_i=o_\PP(1),
	\label{Lun}
\end{align}
where assumption (A7) was applied, with $\widetilde V_i$ a random variable with values in interval $[-2\max(\tau, 1-\tau), 2\max(\tau, 1-\tau)]$. Hence,  $A_n({\cal M}_u)$ becomes:
\[
A_n({\cal M}_u)=\frac{1}{n_{\cal V}} \sum_{i\in {\cal V} } \big[ \rho_\tau(Y_i-\eX^\top_{i,{\cal M}_u} \eb^*_{{\cal M}_u})- \rho_\tau(\varepsilon_i)\big] +o_\PP(1)=\liminf_{n \rightarrow\infty} k_n+a_n >0.
\]

\hspace*{\fill}$\blacksquare$ \\


\end{document}